\documentclass[aps,pra,twocolumn,showpacs,superscriptaddress,reprint]{revtex4}
\usepackage{amsfonts}
\usepackage[utf8]{inputenc}
\usepackage[english]{babel}
\usepackage{amsmath}
\usepackage{amsthm}
\usepackage{amssymb}
\usepackage{newlfont}
\usepackage{graphicx}
\usepackage{epstopdf}
\usepackage{appendix}
\usepackage{hyperref}
\usepackage{textcomp}
\usepackage{multirow}
\usepackage{color}
\usepackage{epsfig}
\usepackage{bm}
\usepackage{braket}
\usepackage{subfigure}
\usepackage{placeins}
\usepackage[normalem]{ulem}

\begin{document}

\author{Ishfaq Ahmad Bhat}
\affiliation{Department of Physics, Pondicherry University, Pondicherry 605014, India}
\author{S. Sivaprakasam}
\affiliation{Department of Physics, Pondicherry University, Pondicherry 605014, India}
\author{Boris A. Malomed}
\affiliation{Department of Physical Electronics, School of Electrical Engineering,
Faculty of Engineering, and Center for Light-Matter Interaction, Tel Aviv
University, Tel Aviv 69978, Israel}
\affiliation{Instituto de Alta Investigaci\'{o}n, Universidad de Tarapac\'{a}, Casilla
7D, Arica, Chile}
\title{ Modulational instability and soliton generation in chiral
Bose-Einstein condensates with \textit{zero-energy} nonlinearity}

\begin{abstract}
By means of analytical and numerical methods, we address the modulational
instability (MI) in chiral condensates governed by the Gross-Pitaevskii
equation including the current nonlinearity. The analysis shows that this
nonlinearity partly suppresses off the MI driven by the cubic self-focusing,
although the current nonlinearity is not represented in the system's energy
(although it modifies the momentum), hence it may be considered as \textit{%
zero-energy nonlinearity}. Direct simulations demonstrate generation of
trains of stochastically interacting chiral solitons by MI. In the
ring-shaped setup, the MI creates a single traveling solitary wave. The sign
of the current nonlinearity determines the direction of propagation of the
emerging solitons.
\end{abstract}

\maketitle

\section{Introduction}

Since the creation of the Bose-Einstein condensates (BECs) in 1995, they
became versatile testbeds for the study of various physical phenomena in
quantum states of matter \cite{Greiner, Bloch}. One of striking properties
of the condensates is their ability to emulate physics of charged particles
under the action of magnetic fields \cite{Lin1, Dalibard, Goldman} through
engineering of synthetic gauge fields in such charge-neutral ultracold
atomic gases.

Synthetic gauge fields can be introduced by means of rapid rotation of the
condensate \cite{Matthews, Madison}, optical coupling between internal
states of atoms \cite{Juzeliuas1, Juzeliuas2, Lin}, laser-assisted tunneling 
\cite{Miyake2013, Aidelsburger}, and Floquet engineering \cite{Parker2013}.
The nature of these gauge fields is essentially static, as parameters of
field-inducing laser beams, including their intensity and phase gradients,
cannot reproduce the time dependence of the Maxwell's equations. Dynamical
gauge fields, affected by nonlinear feedback from matter, to which the
fields are coupled, are required to emulate a full time-dependent field
theory. A number of schemes \cite{Banerjee, Zohar, Tagliacozzo, Greschner,
Dong, Ballantine} have been proposed for creating dynamical gauge fields
with ultra-cold atoms, including the ones which give rise to
density-dependent gauge potentials and current nonlinearities \cite%
{Edmonds1, Zheng}. Following these proposals, such dynamical gauge fields
have been experimentally realized recently \cite{Martinez2016, Clark2018,
Gorg2019}.

The chiral condensates, interacting with gauge potentials, enrich physics of
atomic and nonlinear systems, thus drawing much interest in experimental and
theoretical studies. The presence of density-dependent gauge fields makes it
possible to create anyonic structures \cite{Keilmann} and chiral solitons 
\cite{Edmonds1, Aglietti}, which opens new perspectives for quantum
simulations. The chiral solitons move unidirectionally, the selected
direction being determined by the current nonlinearity. The chiral solitons
were considered for emulation of quantum time crystals (first envisaged by
Wilczek in 2012 \cite{Wilczek}) in circumferentially confined condensates
with density-dependent gauge potentials \cite{Ohberg,
Ohberg2020comment,ohberg2020response}. However, this proposal was disputed
since, in the thermodynamic limit of the latter setting, the lowest-energy
ground state is realized by a static soliton, hence a genuine time crystal
is not feasible \cite{PhysRevResearch.2.032038, syrwid2020response}.

The chiral condensates have also been studied in the presence of persistent
currents \cite{Edmonds1} and collective excitations \cite{Edmonds2} in them.
The evolution of the excitations is affected by the current nonlinearity,
leading to irregular dynamics in a strongly nonlinear regime and related
violation of the Kohn's theorem. The current nonlinearities in chiral
condensates, in addition to being responsible for nonintegrable collision
dynamics in soliton pairs \cite{Dingwall}, help to maintain rich dynamics in
trapped condensates \cite{Saleh}.

In this paper, we study effects of the current nonlinearity on modulational
instability (MI) (alias Benjamin-Feir instability \cite{Benjamin}) of the
chiral BEC. It is well known that MI is a natural precursor to the formation
of solitonic coherent structures, as a result of the interplay between
intrinsic self-interaction of the medium and diffraction or dispersion, and
has been studied in diverse physical settings theoretically \cite{Benjamin,
Agrawal, Mithun_2020} and experimentally \cite{Everitt, Nguyen,
sanz2019interaction}. The MI strongly depends on the nature of two-body
interactions in single-component condensates, where it occurs only in the
case of self-attraction \cite{Theocharis, Salasnich}. However, the
intercomponent interactions make MI\ scenarios more diverse in
multicomponent condensates. In particular, binary condensates with repulsive
interactions are modulationally unstable under the condition of
immiscibility \cite{Goldstein, Kasamatsu1, Kasamatsu2}. Static gauge fields
which impose spin-orbit coupling make condensates still more vulnerable to
MI \cite{Ishfaq, Mithun}. Further, helicoidal gauge potentials break the MI
symmetry and thus strongly modify patterns of instability regions and gain
in the underlying parameter space \cite{Li}. In this connection, we
investigate the effect of density-dependent gauge potentials on the MI and
subsequent generation of solitons. Additionally, we consider the solitons in
a circumferentially confined condensate in a moving reference frame. A
specific peculiarity of the system is that the gauge potential is
represented in the respective Gross-Pitaevskii (GP) equation by a current
nonlinearity, which, however, \emph{is not} represented by any term in the
system's energy. Thus, this term may be identified as an example of \textit{%
zero-energy nonlinearity}. To the best of our knowledge, effects of such
terms on the MI were not studied before.

The subsequent material is organized as follows. Sec. \ref{sec:model}
introduces the model and the corresponding GP equation including the
density-dependent gauge potential. In Sec. \ref{sec:mi} the dispersion
relation produced by the linear MI analysis is derived and discussed. Sec. %
\ref{sec:numerics} reports results of numerical simulations of the system
under the consideration. The work is concluded by Sec. \ref{sec:conclusions}.

\section{The model}

\label{sec:model} We consider a condensate of two-level atoms with the Rabi
coupling imposed by an incident laser beam, as described by the following
mean-field Hamiltonian \cite{Goldman, Edmonds1}:

\begin{equation}
\begin{split}
\mathcal{\hat{H}}& =\bigg(\frac{\mathbf{\hat{p}}^{2}}{2m}+V(\mathbf{r})\bigg)%
\mathbf{\check{I}}+ \\
& 
\begin{pmatrix}
g_{11}\lvert \Psi _{1}\rvert ^{2}+g_{12}\lvert \Psi _{2}\rvert ^{2} & \frac{%
\hbar \Omega _{r}}{2}e^{-i\phi (\mathbf{r})} \\ 
\frac{\hbar \Omega _{r}}{2}e^{i\phi (\mathbf{r})} & g_{22}\lvert \Psi
_{2}\rvert ^{2}+g_{12}\lvert \Psi _{1}\rvert ^{2}%
\end{pmatrix}%
\end{split}
\label{eq:hamilton}
\end{equation}%
where $\mathbf{\hat{p}}$ is the momentum operator, $V(\mathbf{r})$ is the
trapping potential, $\mathbf{\check{I}}$ is the unity matrix, $\Omega _{r}$
is the Rabi-coupling strength, $\phi (\mathbf{r})$ is a spatially varying
phase of the coupling beam, and $g_{\mu \nu }=\left( 2\pi \hbar
^{2}/m\right) a_{\mu \nu }$ are the mean-field interaction strengths,
proportional to respective scattering lengths, $a_{\mu \nu }$, of collisions
between atoms in internal states $\mu $ and $\nu $ ($\mu ,\nu =1,2$). At the
mean-field level, atomic collisions result in effective density-dependent
detunings of the atomic states. In the case of dilute condensates, when the
Rabi-coupling energy dominates over the collisional detunings, the
perturbative approach leads to the following density-dependent gauge
potential \cite{Edmonds1}: 
\begin{equation}
\mathbf{A}=\mathbf{A}_{0}+n~\frac{g_{11}-g_{22}}{8\Omega _{r}}\nabla \phi (%
\mathbf{r})  \label{eq:ddp}
\end{equation}%
where $\mathbf{A}_{0}=-\left( \hbar /2\right) \nabla \phi (\mathbf{r})$
represents the single-particle vector potential, and $n$ is the density of
the condensate. By projecting the coupled two-level atom onto a single
dressed state, one arrives at the following mean-field GP equation governing
the dynamics of chiral condensates: \cite{Aglietti, Edmonds1, Edmonds2,
Saleh}: 
\begin{equation}
i\hbar \frac{\partial \Psi }{\partial t}=\bigg[\frac{(\mathbf{\hat{p}}-%
\mathbf{A)^{2}}}{2m}+W(\mathbf{r})+V(\mathbf{r})+\mathfrak{g}|\Psi |^{2}+%
\mathbf{a_{1}}\cdot \mathbf{J}(\Psi ,\Psi ^{\ast })\bigg]\Psi  \label{eq:1}
\end{equation}%
where $\Psi $ is the wave function of the dressed state. Further, $W(\mathbf{%
r})=\lvert A_{0}\rvert ^{2}/\left( 2m\right) $, together with the trapping
potential, $V(\mathbf{r})$, defines the scalar potential, $\mathbf{a_{1}}%
=\left( g_{11}-g_{22}\right) /\left( 8\Omega _{r}\right) \nabla \phi (%
\mathbf{r})$ is the strength of the density-dependent vectorial gauge
potential, and $\mathfrak{g}=(1/4)(g_{11}+g_{22}+2g_{12})$ determines the
effective interaction in the dressed-state picture. The unconventional
nonlinearity in the chiral condensates manifests itself in Eq. \eqref{eq:1}
through the current, 
\begin{equation}
\mathbf{J}(\Psi ,\Psi ^{\ast })=\frac{\hbar }{2im}\bigg[\Psi \bigg(\nabla +%
\frac{i}{\hbar }\mathbf{A}\bigg)\Psi ^{\ast }-\Psi ^{\ast }\bigg(\nabla -%
\frac{i}{\hbar }\mathbf{A}\bigg)\Psi \bigg]  \label{eq:current}
\end{equation}%
Defining $\phi (x)=k_{l}x$, where $k_{l}$ is the wavenumber of the incident
laser beam, following the usual procedure of the dimensional reduction \cite%
{Dingwall}, and applying the nonlinear phase transformation, 
\begin{equation}
\Psi (x,t)=\psi (x,t)~\exp \left[ -i\frac{\phi (x)}{2}+i\frac{a_{1}}{\hbar }%
\int_{-\infty }^{x}dx^{\prime }n(x^{\prime },t)-i\frac{Wt}{\hbar }\right] 
\text{,}  \label{eq:nlt}
\end{equation}%
Eq. (\ref{eq:1}) is cast in the form of the following one-dimensional (1D)
GP equation: 
\begin{equation}
i\hbar \frac{\partial }{\partial t}\psi =\bigg(-\frac{\hbar ^{2}}{2m}\frac{%
\partial ^{2}}{\partial x^{2}}+V(x)+g|\psi |^{2}-2\alpha J(x)\bigg)\psi
\label{eq:2}
\end{equation}%
where $V(x)=\left( 1/2\right) m\omega _{x}^{2}x^{2}$ is the 1D trapping
potential with axial trapping frequency $\omega _{x}$, $g\equiv \mathfrak{g}%
/2\pi a_{\perp }^{2}$ is the usual coefficient of the cubic nonlinearity,
and $a_{\perp }=\sqrt{\hbar /\left( m\omega _{\perp }\right) }$ is the
harmonic-oscillator length of the transverse confining potential. In Eq. (%
\ref{eq:2}) $\alpha =\left( 2\pi a_{\perp }^{2}\right) ^{-1}\left[
k_{l}(g_{11}-g_{22})/\left( 8\Omega _{r}\right) \right] $ is the strength of
the unconventional current nonlinearity, which involves the current density, 
$J(x)=\left( i\hbar /2m\right) (\psi \partial _{x}\psi ^{\ast }-\psi ^{\ast
}\partial _{x}\psi )$. Spatiotemporal rescaling,

\begin{equation}
t^{\prime }=\omega _{\perp }t,~~x^{\prime }=\frac{x}{a_{\perp }},~\psi
^{\prime }=\sqrt{a_{\perp }}\psi ,  \label{eq:4}
\end{equation}%
casts Eq. \eqref{eq:2} in the normalized form: 
\begin{equation}
i\partial _{t}\psi =\bigg(-\frac{1}{2}\partial _{xx}+\sigma |\psi
|^{2}-2\rho J(x)\bigg)\psi .  \label{eq:3}
\end{equation}%
In Eq. \eqref{eq:3}, the current density is now defined as 
\begin{equation}
J(x)=\left( i/2\right) (\psi \partial _{x}\psi ^{\ast }-\psi ^{\ast
}\partial _{x}\psi ),  \label{J}
\end{equation}%
and the axial potential, which is $V(x)$ in Eq. (\ref{eq:2}), is dropped, as
the following analysis deals with the dynamics in free space. The
generalized GP equation (\ref{eq:3}), with scaled interaction parameters $%
\sigma =\left( a_{11}+a_{22}+2a_{12}\right) /\left( 2a_{\perp }\right) $ and 
$\rho =\left( \hbar k_{l}/4m\Omega _{r}\right) (a_{11}-a_{22})/a_{\perp
}^{2} $, is used below for the consideration of MI and formation of chiral
solitons. Note that solutions of Eq. (\ref{eq:3}) with $\sigma <0$ for
quiescent solitons, with an arbitrary real chemical potential, $\mu <0$, are
precisely the same as in the case of $\rho =0$:%
\begin{equation}
\psi _{\mathrm{sol}}=e^{-i\mu t}\sqrt{2\mu /\sigma }\mathrm{sech}\left( 
\sqrt{-2\mu }x\right) .  \label{soliton}
\end{equation}

Equations (\ref{eq:3}) and (\ref{J}) conserve the integral norm
(proportional to the total number of particles), $N=\int_{-\infty }^{+\infty
}dx|\psi |^{2}$, energy, 
\begin{equation}
E=-\frac{1}{2}\int_{-\infty }^{+\infty }dx\left( \left\vert \frac{\partial
\psi }{\partial x}\right\vert ^{2}+\sigma |\psi |^{4}\right) ,  \label{E}
\end{equation}%
and momentum, 
\begin{equation}
P=-\int_{-\infty }^{+\infty }dx\left( J(x)+\rho |\psi |^{4}\right)  \label{P}
\end{equation}%
\cite{Dingwall, Jackiw}. Unlike the underlying chiral GP equation (\ref{eq:1}%
), the transformed equation (\ref{eq:3}) cannot be written in the Lagrangian
form. It cannot be written in the Hamiltonian form either, but,
nevertheless, energy $E$, defined by Eq. (\ref{E}), is its dynamical
invariant. It is worthy to note that the energy conservation holds in spite
of the fact $E$ does not include any term $\sim \rho $ (in fact, $E$ is the
same as for the usual non-chiral GP equation), i.e., the current term may be
identified as \textit{zero-energy nonlinearity} [although it affects
expression (\ref{P}) for the conserved momentum]. The conservation of $E$
can be readily verified by the straightforward calculation of $dE/dt$,
substituting $\partial \psi /\partial t$ and $\partial \psi ^{\ast
}/\partial t$ in the integrand by what is produced by Eq. (\ref{eq:3}). As a
result, the terms $\sim \rho $ amount to expressions in the form of full $x$%
-derivatives, hence their contribution to the integral expression for $dE/dt$
identically vanishes for all localized states, the result being $dE/dt=0$.

Qualitatively, a term in conservative equations which does not affect the
energy may be compared to the Lorentz force acting on a charged particle in
magnetic field, or the Coriolis force for a body moving on a rotating
sphere. However, such an analogy is not an accurate one, because the Lorentz
and Coriolis forces, although they are not represented in the particle's
Hamiltonian, appear in the Lagrangian, while, as mentioned above, the term $%
\sim \rho $ in Eq. (\ref{eq:3}) cannot be derived from a Lagrangian. In
terms of models governed by GP-like equations, somewhat similar effects are
produced by the stimulated Raman scattering (SRS)\ in fiber optics \cite%
{Raman} and its \textit{pseudo-SRS} counterpart in the system of interacting
high- and low-frequency waves (the Zakharov system) \cite{Gromov}, as well
as by the nonlinear Landau damping in plasmas \cite{Landau}. However, in
those cases the non-Lagrangian terms conserve only the norm of the wave
fields, while causing the dissipation of the energy and momentum (in
particular, self-decelaration of solitons \cite{Raman}), therefore the
latter analogy is not accurate either.

On the other hand, the current nonlinearity in Eq. (\ref{eq:3}) makes the
total momentum, as given by Eq. (\ref{P}), different from the standard
expression for the non-chiral GP equation. In this connection, it is also
relevant to mention that, unlike the usual GP equation, Eq. (\ref{eq:3}) is
not invariant with respect to the Galilean transform: the substitution of 
\begin{equation}
\tilde{x}=x-ut,~\psi =\exp \left( iux-iu^{2}t/2\right) \tilde{\psi}\left( 
\tilde{x},t\right) ,  \label{Galileo}
\end{equation}%
transforms Eq. (\ref{eq:3}) not into the equation of the same form for wave
function $\tilde{\psi}\left( \tilde{x},t\right) $, written in the reference
frame moving with arbitrary velocity $u$, but into one with $\sigma $
replaced by%
\begin{equation}
\tilde{\sigma}=\sigma -2\rho u.  \label{sigma}
\end{equation}%
Note that taking $\rho u>\sigma /2$ replaces $\sigma >0$ (self-repulsion) by 
$\tilde{\sigma}<0$ (self-attraction), and vice versa, taking $\rho u<\sigma
/2$, in the case of $\sigma <0$.

According to Eqs. (\ref{Galileo}) and (\ref{sigma}), the solution of Eq. (%
\ref{eq:3}) for a soliton moving with velocity $u$ is obtained from the
quiescent one (\ref{soliton}) \cite{Dingwall}:%
\begin{equation}
\begin{split}
\psi _{\mathrm{sol}}& =\exp \left( iux-i\left( \mu +u^{2}/2\right) t\right) 
\sqrt{2\mu /\left( \sigma -2\rho u\right) } \\
& \times \mathrm{sech}\left( \sqrt{-2\mu }\left( x-ut\right) \right) .
\end{split}
\label{u-sol}
\end{equation}%
Obviously, this solution exists for $\mu <0$ and $2\rho u-\sigma >0$. In the
framework of the non-integrable GP equation (\ref{eq:3}), collisions between
solitons moving at different velocities were studied in Ref. \cite{Dingwall}%
, revealing various inelastic outcomes of the collisions.

Lastly, as concerns dissipative effects, parameters of the loss, chiefly
induced by collisions of atoms from the condensate with ones belonging to
the thermal component of the gas, were calculated, in particular, in Ref. 
\cite{Choi}. Estimates based on these values easily demonstrate that the
dissipation is not an essential factor for experimentally relevant times.

\section{The linear-stability analysis (LSA)}

\label{sec:mi} We examine the MI via linear instability analysis (LSA) of
the spatially uniform (alias continuous-wave, CW) state of the chiral
condensate in the absence of the external potential, $V(x)=0$. The
respective linearized Bogoliubov-de Gennes equations for small perturbations
about the CW wave function were derived in Refs. \cite{Edmonds2, Dingwall},
though the MI spectrum has not been derived, and is addressed here. The
usual CW solution to Eq. (\ref{eq:3}) is 
\begin{equation}
\psi (x,t)=\sqrt{n}\exp \left( iKx-i\mu _{\mathrm{CW}}t\right) ,
\label{eq:5a}
\end{equation}%
with arbitrary density, $n=|\psi |^{2}$, real wavenumber $K$, and chemical
potential $\mu _{\mathrm{CW}}={\sigma n-2\rho K}$. It is more convenient to
develop the MI analysis in the reference frame moving at velocity $u=K$,
applying the Galilean transformation provided by Eq. (\ref{Galileo}). It
removes term $Kx$ from the CW phase, replacing $\sigma $ by 
\begin{equation}
\tilde{\sigma}=\sigma -2\rho K,  \label{tilde}
\end{equation}%
as per Eq. (\ref{sigma}).

We perturb the CW state~\eqref{eq:5a}, replacing it by (in the moving
reference frame, if $K\neq 0$) 
\begin{equation}
\psi =(\sqrt{n}+\delta \psi \left( x,t\right) )\exp \left( -i\mu _{\mathrm{CW%
}}t\right) ,  \label{eq:5}
\end{equation}%
which results in the following linearized equation for the small
perturbation, $\delta \psi (x,t)$: 
\begin{equation}
i\partial _{t}(\delta \psi )=-\frac{1}{2}\partial _{xx}^{2}(\delta \psi )+n%
\tilde{\sigma}(\delta \psi +\delta \psi ^{\ast })+in\rho \left( \partial
_{x}\delta \psi -\partial _{x}\delta \psi ^{\ast }\right)  \label{eq:6}
\end{equation}%
where $\ast $ stands for the complex conjugate, and $x$ is written instead
of $\tilde{x}$, to simplify the notation, if the moving reference frame is
used. We look for eigenmodes of the perturbations in the usual plane-wave
form, $\delta \psi =a\cos (kx-\Omega t)+ib\sin (kx-\Omega t)$, with real
wavenumber $k$ and complex eigenfrequency, $\Omega $, assuming that the
actual system's size is much greater than the healing length which
determines a characteristic MI\ length scale. The substitution of this in
Eqs. \eqref{eq:6} and (\ref{J}) results in the following dispersion
equation: 
\begin{equation}
\Omega ^{2}+2n\rho k~\Omega -n\tilde{\sigma}k^{2}-\frac{k^{4}}{4}=0,
\label{eq:7}
\end{equation}%
which yields two branches of the $\Omega (k)$ dependence: 
\begin{equation}
\Omega _{\pm }=-n\rho k\pm \sqrt{\frac{k^{4}}{4}+k^{2}n(n\rho ^{2}+\tilde{%
\sigma})}  \label{eq:8}
\end{equation}%
Equation \eqref{eq:8} is the Bogoliubov dispersion relation \cite{Bogolyubov}
for the propagation of small perturbations on top of the CW background. The
expression on the right-hand side of Eq. \eqref{eq:8} may be positive,
negative or complex, depending on the signs and magnitudes of the
interaction parameters. The CW solutions are stable if $\Omega _{\pm }(k)$
are real for all real $k$; otherwise, the instability gain is defined as $%
\xi \equiv \lvert \text{Im}(\Omega _{\pm })\rvert $. For $\rho =0$ (no
current nonlinearity), the above consideration reproduces the well-known
results for BEC with the cubic self-attraction ($\sigma <0$), yielding the
MI in the wavenumber band $0<k<2\sqrt{n\lvert \sigma \rvert }$, with the
maximum MI gain, $\xi _{\max }=n\lvert \sigma \rvert $, attained at $k_{\max
}=\sqrt{2n\lvert \sigma \rvert }$ \cite{Benjamin, Theocharis}.

\begin{figure}[tbh]
\begin{center}
\includegraphics[width=0.5\textwidth]{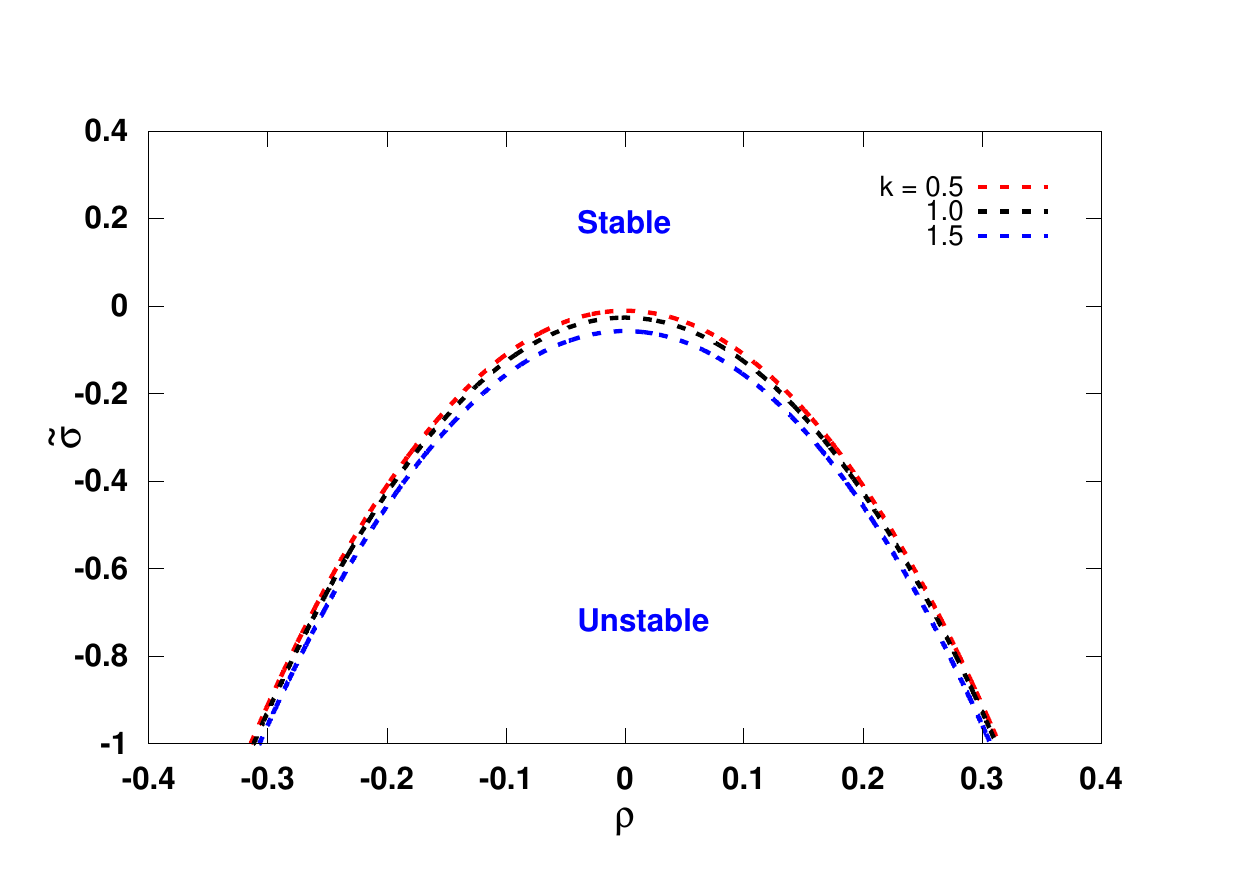}
\end{center}
\caption{(Color online) The MI\ stability diagram, determined by Eq. (%
\protect\ref{eq:8}) for $n=10$ and different values of $k$ in the $\left( 
\tilde{\protect\sigma},\protect\rho \right) $ plane: as it follows Eq. (%
\protect\ref{eq:8}), the CW state is modulationally stable for $\protect\rho %
^{2}\geq \left( k^{2}-4\tilde{\protect\sigma}n\right) /\left( 4n^{2}\right) $%
. Note the symmetry of the curves about $\protect\rho =0$. }
\label{fig:1}
\end{figure}

The effect of the current nonlinearity, quantified by coefficient $\rho $,
on the MI can be understood in the framework of Eq. \eqref{eq:8}. It is
evident that the MI gain is essentially the same for both signs of the
current nonlinearity, $\rho $. Further, such chiral condensates may be
modulationally unstable only for the attractive sign of the two-body
interactions, \textit{i.e.}, $\tilde{\sigma}<0$. The current-nonlinearity's
strength, $\rho $, controls the MI region and gain for a fixed value of $%
\tilde{\sigma}$, as shown in Fig. \ref{fig:1}. It is found from Eq. %
\eqref{eq:8} that the system is subject to the MI in the wavenumber range of 
$0<k<2\sqrt{n\lvert \tilde{\sigma}\rvert -n^{2}\rho ^{2}}$, provided that 
\begin{equation}
\rho <\sqrt{\lvert \tilde{\sigma}\rvert /n}.  \label{rho<}
\end{equation}%
For $\rho \geq \sqrt{\lvert \tilde{\sigma}\rvert /n}$, the CW state is
modulationally stable even for attractive two-body interactions, thereby
revealing the stabilizing effect of the current nonlinearity in the chiral
condensates. Further, substituting definition (\ref{tilde}) in Eq. (\ref%
{rho<}), the parameter region in which the MI takes place for $\rho K\neq 0$
amounts to the following interval of the current-nonlinearity's strength:%
\begin{equation}
K-\sqrt{K^{2}-\sigma n}<n\rho <K+\sqrt{K^{2}-\sigma n},  \label{KK}
\end{equation}%
which exists provided that the cubic-nonlinearity coefficient satisfies
condition%
\begin{equation}
\sigma n<K^{2}.  \label{K^2}
\end{equation}%
In particular, this condition holds for all $\sigma <0$, while $K\neq 0$ may
impose the MI even in the case when the cubic term in Eq. (\ref{eq:3}) is
self-defocusing, with $\sigma >0$. 

\begin{figure}[tbp]
\begin{center}
\includegraphics[width=0.5\textwidth]{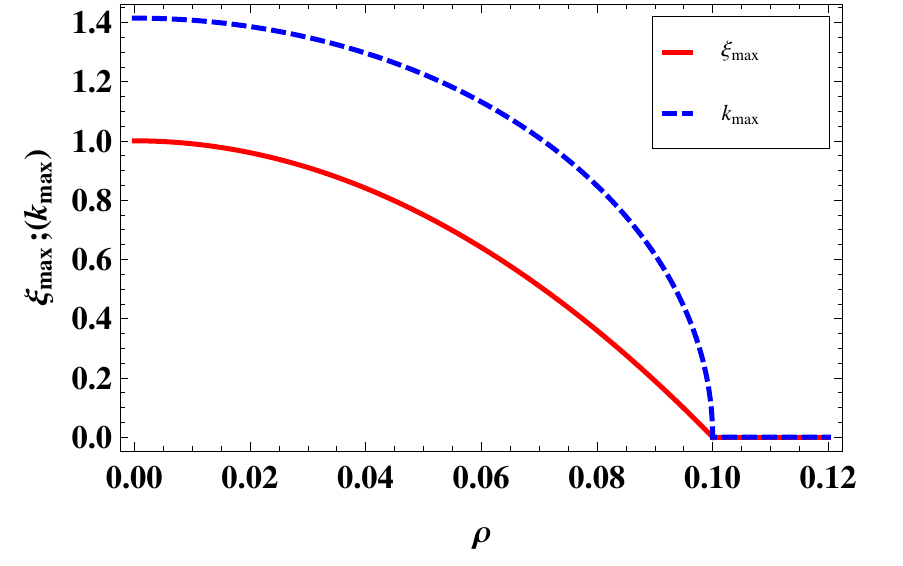}
\end{center}
\caption{(Color online) Variation of $\protect\xi _{\max }$ and $k_{\max }$
as a function of the current-nonlinearity's strength, $\protect\rho $, for $%
n=10~\text{and}~\tilde{\protect\sigma}=-0.1$, as per Eqs. (\protect\ref{eq:9}%
) and Eq. (\protect\ref{eq:10}). The MI boundary, $\protect\rho =0.1$, is
given by Eq. (\protect\ref{rho<}).}
\label{fig:2}
\end{figure}

In combination with the Feshbach-resonance techniques \cite{Blatt, Inouye,
Cornish}, which makes it possible to adjust the value of $\sigma $, the
chiral-interaction strength, $\rho ,$ can be used to tune the MI gain,
including suppression of the instability. This possibility can be adequately
described in terms of the change of the maximum MI\ gain $\xi _{\max }$ at $%
k=k_{\max }$, following the variation of $\rho $. It follows from Eq. %
\eqref{eq:8} that, for $\rho <\sqrt{\lvert \tilde{\sigma}\rvert /n}$, the
largest MI gain, 
\begin{equation}
\xi _{\max }=n(|\tilde{\sigma}|-n\rho ^{2})  \label{eq:9}
\end{equation}%
is attained at 
\begin{equation}
k_{\max }=\pm \sqrt{2n(|\tilde{\sigma}|-n\rho ^{2})}.  \label{eq:10}
\end{equation}%
It is seen from Eqs. \eqref{eq:9} and \eqref{eq:10} that both $\xi _{\max }$
and $k_{\max }$ decrease monotonically with the increase of $\rho $, up to $%
\rho =\sqrt{\lvert \tilde{\sigma}\rvert /n}$, as shown in Fig. \ref{fig:2}.
The value $\xi _{\max }$ vanishes at $\rho ^{2}=|\tilde{\sigma}|/n$, which
is the above-mentioned MI boundary, given by Eq. (\ref{rho<}). 

\begin{figure}[tbp]
\begin{center}
\includegraphics[width=0.4\textwidth]{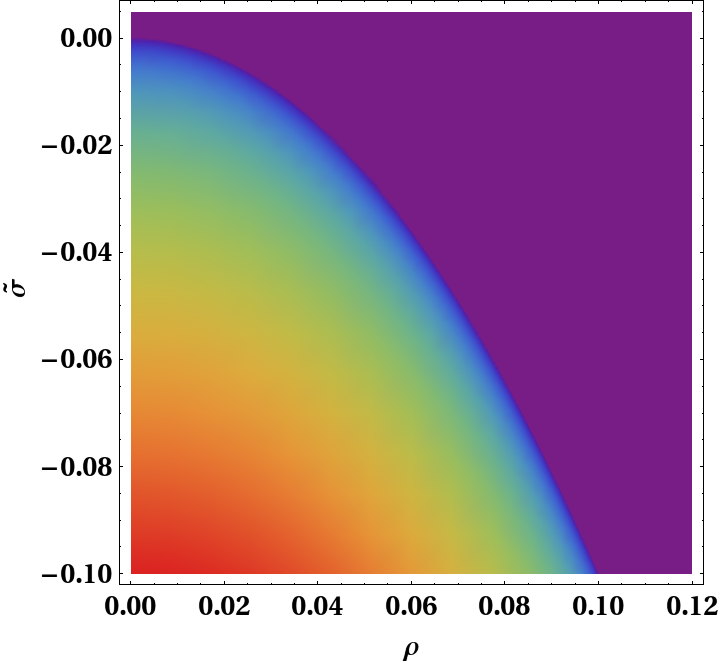} %
\includegraphics[width=0.9 cm, height= 6.6 cm]{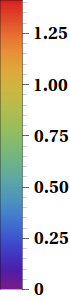}
\end{center}
\caption{(Color online) The perturbation wavenumber, $k_{\max }$, at which
the MI gain attains its maximum, as a function of the strengths of the
two-body (cubic) attraction, $\tilde{\protect\sigma} $, and the current
nonlinearity, $\protect\rho $, for $n=10$.}
\label{fig:3}
\end{figure}

A plot of the variation of $k_{\max }$ with the change of the strength of
the nonlinearity coefficients, $\rho $ and $\tilde{\sigma} $, is presented,
in the form of a heatmap, in Fig. \ref{fig:3}, as per Eq. \eqref{eq:10}.
This plot also shows that the addition of the current nonlinearity results
in the stabilization of the condensate against the modulational
perturbations. 

\begin{figure}[tbh]
\includegraphics[width=0.5\textwidth]{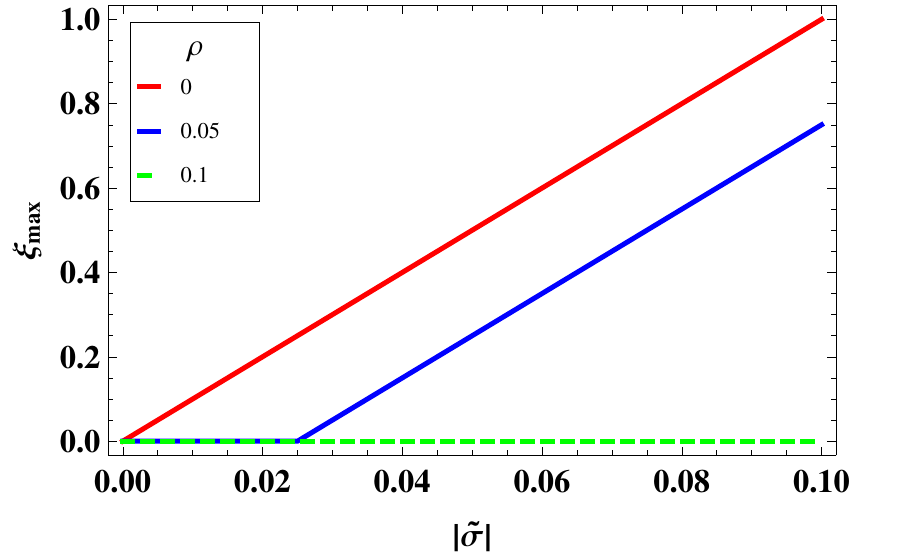}
\caption{(Color online) Variation of $\protect\xi _{\max }$ as a function of
the strength of the two-body interaction, $\lvert \tilde{\protect\sigma}%
\rvert $, for three fixed values of the current-nonlinearity's strength, $%
\protect\rho $ and $n=10$.}
\label{fig:4}
\end{figure}

Next we study the variation of MI gain with respect to the strength of the
two-body attraction, $\tilde{\sigma}$, at a fixed value of the
current-nonlinearity's strength, $\rho $. As mentioned above, the system is
modulationally unstable if and only if the two-body interaction is
attractive and condition (\ref{rho<}) holds. The maximum MI gain, $\xi
_{\max }$, as given by Eq. \eqref{eq:9}, is plotted in Fig. \ref{fig:4}.
Accordingly, the dependence of $\xi _{\max }$ on $\tilde{\sigma}$ is linear,
with the slope determined by the CW density $n$. 

\section{Numerical results}

\label{sec:numerics} Proceeding to the numerical analysis, we employed a
sixth-order Runge-Kutta scheme for simulations of the GP equation~(\ref{eq:3}%
), cf. Refs. \cite{sarafyan1972improved,PhysRevA.51.1382}. The
kinetic-energy term was dealt with by dint of the fast Fourier transform,
while the spatial derivative in the current-nonlinearity part we
approximated by the fourth-order central-difference formula. In the
simulations, we fixed the timestep, $\Delta t=0.0001$, the domain size, $%
L=100$, and the spatial mesh size, $\Delta x=L/N$ with $N=2048$, unless
mentioned otherwise. 
\begin{figure}[tbp]
\includegraphics[width=0.5\textwidth]{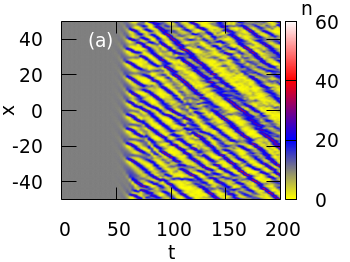} \includegraphics[width=0.5%
\textwidth]{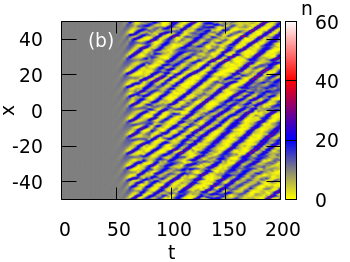}
\caption{(Color online) The evolution of the condensate density due to the
development of the MI for the cubic-attraction strength $\protect\sigma %
=-0.1 $ with (a) $\protect\rho =0.08$, and (b) $\protect\rho =-0.08$.}
\label{fig:5}
\end{figure}
\begin{figure}[tbp]
\includegraphics[width=0.5\textwidth]{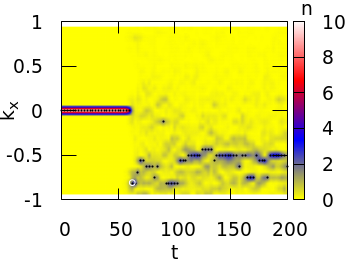}
\caption{(Color online) The evolution of the condensate density in the $k$%
-space for the cubic-attraction and current-nonlinearity's strengths $%
\protect\sigma =-0.1$ and $\protect\rho =0.08$. The circle at $t=62$
indicates the wavenumber, $k_{\max }\approx 0.81$, at which the MI-induced
mode features the maximum of $n(k_{x})$, see the text.}
\label{fig:5a}
\end{figure}

The initial condition was taken as the CW background with uniform density, $%
n=10$, and $K=0$ [see Eq. (\ref{eq:5a})], seeded by weak random
perturbations. It is clearly seen from Fig. \ref{fig:2} that the system is
modulationally unstable for $\sigma =-0.1$, provided that $\rho <0.1$.
Figure \ref{fig:5}(a) shows the spatiotemporal evolution of the initially
perturbed CW density for parameters $\sigma =-0.1,~\rho =0.08$. It shows
generation of a train of chiral solitons at $t\geq 50$, propagating in the
negative $x$-direction. The unidirectional propagation is a signature of the
chiral solitons, see Eq. (\ref{u-sol}). Further, solitons belonging to the
train collide inelastically due to the nonintegrability of the model \cite%
{Dingwall}. The direction of motion of the solitons is determined by the
sign of $\rho $, while the occurrence of the MI is independent of this sign,
as shown in the previous section. To confirm this expectation, in Fig. \ref%
{fig:5}(b) we set $\rho =-0.08$ for the same $\sigma =-0.1$. This simulation
gives rise to a soliton train, with the same structure, but propagating in
the positive $x$-direction. The choice of the propagation direction by the
generated solitons can be understood from the $k$-space density, $n(k_{x})$,
shown in Fig. \ref{fig:5a}, where modes with $k_{x}<0$ are chiefly excited.

The wavenumber corresponding to the maximum of $n\left( k_{x}\right) $ in
the spontaneously excited mode is $k_{x}\approx 0.81$. Note that it is close
to $k_{\max }=0.84$ predicted by the LSA. Further, values $k_{\max }=0.84$
and $\xi _{\max }=0.36$ correspond, severally, to the fastest growing mode
with wavelength $\lambda _{\max }\equiv 2\pi /k_{\max }\approx 7.4$ and
growth time scale $\tau =2\pi /\xi _{\max }\approx 17$. In turn, $\lambda
_{\max }$ determines the number of solitons in the train observed in Fig. %
\ref{fig:5}, as $\mathrm{n}_{\mathrm{sol}}\simeq L/\lambda \approx 14$,
where $L=100$ is the above-mentioned size of the simulation domain. 
\begin{figure}[tbh]
\begin{center}
\includegraphics[scale=1,width=0.49\textwidth]{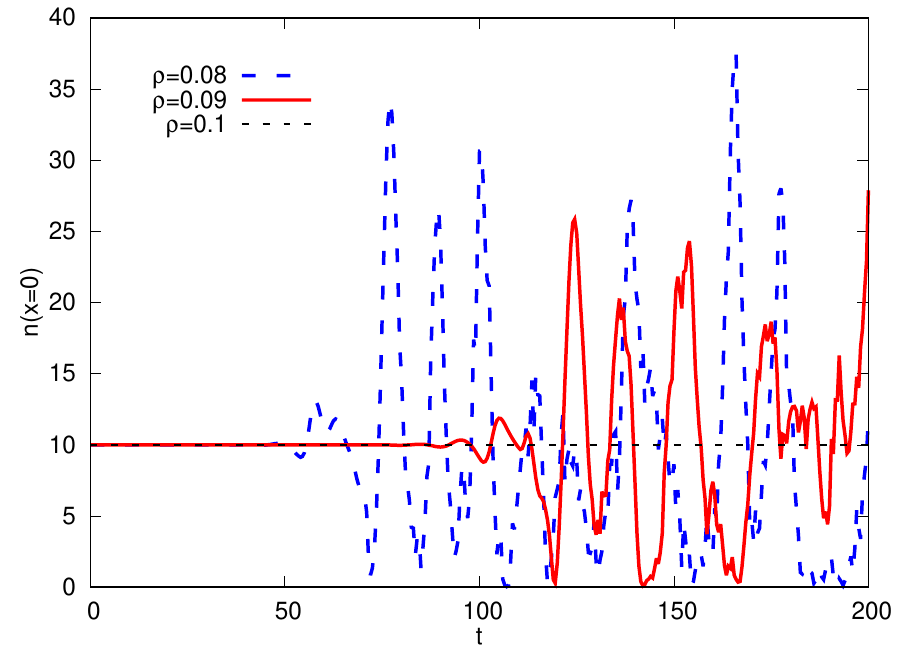}
\end{center}
\caption{(Color online) The evolution of MI at different values of the
current-nonlinearity's strengths, $\protect\rho $, and fixed self-attraction
coefficient, $\protect\sigma =-0.1$, is displayed in terms of the midpoint
density, $n(x=0)$. For $\protect\sigma <0$, the condensate is modulationally
stable at $\protect\rho \geq \protect\sqrt{|\protect\sigma |/n}$, see Eq. (%
\protect\ref{rho<}).}
\label{fig:6}
\end{figure}

We extend the numerical analysis by varying $\rho $ at fixed $\sigma $.
Figure~\ref{fig:6} shows the temporal evolution of the midpoint density, $%
n(x=0)$, for $\rho =(0.1,0.09,0.08)$. The MI condition (\ref{rho<}),
predicted by LSA, is exactly confirmed by the figure. Further, it is seen
that, as $\rho $ increases towards value $\sqrt{|\sigma |/n}$, above which
the MI is suppressed, instability-growth time increases, due to the decrease
in the MI gain, in agreement with Fig.~\ref{fig:2}. Numerical data produced
by these simulations are summarized in Table \ref{tab1}. It is clearly seen
that the findings agree with the predictions of the LSA analysis.%

\begin{center}
\begin{table}[tbh]
\caption{The summary of the MI characteristics produced by the LSA and
numerical simulations for the fixed cubic-attraction coefficient, $\protect%
\sigma =0.1$, fixed CW density, $n=10$, and different values of the
current-nonlinearity's strength, $\protect\rho $. Wavenumber $k_{\max }$ and
the respective wavelength, $\protect\lambda =2\protect\pi /k_{\max }$,
corresponding to the largest MI gain, $\protect\xi _{\max }$, are given by
Eq. (\protect\ref{eq:10}). The MI growth time, $\protect\tau =1/\protect\xi %
_{\max }$, is determined by Eq. (\protect\ref{eq:9}). The number of solitons
in the MI-generated train is accurately approximated by estimate $\mathrm{n}%
_{\mathrm{sol}}\simeq L/\protect\lambda $, see the text.}
\label{tab1}
{\footnotesize \centering
\begin{tabular}{p{1cm}p{1cm}p{1cm}p{1cm}p{1cm}p{1cm}c}
\hline\hline
$\sigma$ & $\rho$ & $k_{max}$ & $\lambda$ & $\tau$ & number of solitons &  \\%
[0.5ex] \hline
\multirow{4}{*}{$ -0.1$} & $0.05$ & $1.22$ & $5.13$ & $8.37$ & $\sim 19$ & 
\\ 
& $0.08$ & $0.84$ & $7.40$ & $17.45$ & $\sim 14$ &  \\ 
& $0.09$ & $0.61$ & $10.19$ & $33.06$ & $\sim 10$ &  \\ 
& $0.1$ & $0$ & $\infty$ & $\infty$ & $~~~~0$ &  \\ \hline\hline
\end{tabular}
}
\end{table}
\end{center}

\begin{figure}[tbh]
\includegraphics[width=0.5\textwidth]{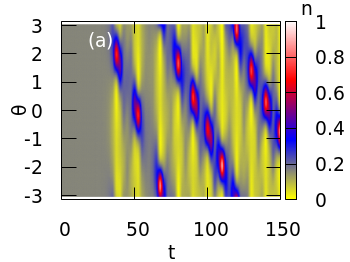} %
\includegraphics[width=0.5\textwidth]{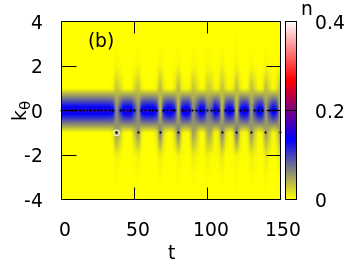} 
\caption{(Color online) Numerically simulated evolution of MI in a
circumferentially confined condensate with $\tilde{g}=-3.6$, $\mathit{a}=%
\frac{\protect\pi }{4}$ and $\protect\int_{0}^{2\protect\pi }d\protect\theta %
|\protect\varphi (\protect\theta ,\protect\tau )|^{2}=1$. The white circle
represents the wave vector corresponding to the initially excited mode.}
\label{fig:8}
\end{figure}

Lastly, we address the 1D setting corresponding to the ring geometry with
radius $R$ and periodic boundary conditions, the accordingly rescaled form
of Eqs. (\ref{eq:3}) and (\ref{J}) being 
\begin{equation}
i\frac{\partial \varphi (\theta ,\tau )}{\partial \tau }=\bigg(-\frac{1}{2}%
\frac{\partial ^{2}}{\partial \theta ^{2}}+\tilde{g}|\varphi |^{2}-2\mathit{a%
}J(\theta )\bigg)\varphi (\theta ,\tau ),  \label{eq:13}
\end{equation}%
with $J(\theta )=(i/2)\text{Im}(\varphi \partial _{\theta }\varphi ^{\ast
}-\varphi ^{\ast }\partial _{\theta }\varphi )$ and wave function $\varphi $
subject to the normalization condition $\int_{0}^{2\pi }d\theta |\varphi
(\theta ,\tau )|^{2}=1$, where $\theta $ is the angular coordinate on the
ring. The accordingly scaled parameters are $\tilde{g}=\left( g-2\alpha
u\right) \left( mR^{2}/\hbar ^{2}\right) $ and $\mathit{a}=\alpha R/\hbar $,
with $\alpha \equiv q(g_{11}-g_{22})/\left( 8\Omega _{r}R\right) $
determined by integer winding number $q$ of the laser beam used to induce
the circumferential gauge potentials. In this case, the CW state (\ref{eq:5}%
), including term $K\theta $ in the phase, can be constructed too, with
integer values of $K$ allowed by the circular boundary condition.

As typical parameter values, we take $\mathit{a}=\pi /4$, $u=-1/4$ and $g=-4$%
, the respective cubic-term coefficient in Eq. (\ref{eq:13}) being $\tilde{g}%
=-3.6$. Note that this value satisfies condition $g<-\pi$, which is
necessary to the transition from the uniform state in the ring to a
soliton-like pattern \cite{Ueda}. For this case, Fig. \ref{fig:8} shows the
nonlinear evolution of the weak perturbations, initiated by the MI, that
ultimately coalesce into a single chiral soliton, which performs circular
motion on the ring. Note that the system does not converge to an equilibrium
state that would be appropriate for the realization of the quantum time
crystal. In addition to the spatial distribution of the density, displayed
in \ref{fig:8}(a), panel \ref{fig:8}(b) shows that the wavenumber
corresponding to the initially excited mode is $|k_{\theta }|=1$. It matches
well to the analytical estimate given by the Eq.~(\ref{eq:10}), even if it
was derived for the infinite system, rather than for the circumference. As
mentioned above, owing to the ring geometry of the condensate, only integer
values are allowed for wavenumbers $k$. Consequently, the condensate
undergoes oscillation in the $k-$space, as observed in Fig.~\ref{fig:8}(b).


\section{Conclusion}

\label{sec:conclusions} We have studied the MI (modulational instability) of
the chiral condensate by means of LSA (linear stability analysis) and direct
simulations. The chirality is introduced by the action of the current
nonlinearity. A peculiar property of this nonlinearity is its \textit{%
zero-energy} character, as, being accounted for by term $\sim \rho $ in the
underlying GP equation (\ref{eq:3}), it is not represented in the respective
energy, given by Eq. (\ref{E}). Nevertheless, the current nonlinearity
strongly affects the MI, tending to suppress it. Direct simulations
demonstrate that, as long as the MI is present, it generates a train of
stochastically interacting chiral solitons in the extended system, or a
single soliton in the ring-shaped one. The MI gain does not depend on the
sign of the current nonlinearity, although the sign determines the direction
of the motion of the generated solitons.

An interesting extension of the work may be the consideration of GP
(Gross-Pitaevskii) equations including the current nonlinearity in a
combination with nonlinear terms that give rise to the onset of the critical
collapse. In the 1D setting, the collapse is induced by the self-focusing
quintic term, which may represent attractive three-body interactions in the
atomic condensate \cite{quintic1,quintic2}. A crude estimate of the virial
type \cite{Fibich} suggests that the current nonlinearity, although it is
not represented in the system's energy, may compete on a par with the
quintic self-attraction, hence it may essentially affect the collapse
dynamics. Further, the same current term can be added, as a quasi-1D one, to
the two-dimensional (2D) GP equation, with the usual cubic self-attraction,
which drives the development of the critical collapse in 2D \cite{Fibich}.
An estimate suggests that, in the 2D setup, the current nonlinearity will be
the strongest term in the collapsing regime, hence its effect may be
crucially important. 

\section{Acknowledgements}

We thank Dr. T. Mithun for useful discussions. The work of B.A.M. was
supported, in part, by the Israel Science Foundation through grant No.
1286/17.

%
\let\itshape\upshape
\bibliographystyle{apsrev4}
\bibliography{reference}

\providecommand{\noopsort}[1]{}\providecommand{\singleletter}[1]{#1}%
\begin{thebibliography}{64}%
\makeatletter
\providecommand \@ifxundefined [1]{%
 \@ifx{#1\undefined}
}%
\providecommand \@ifnum [1]{%
 \ifnum #1\expandafter \@firstoftwo
 \else \expandafter \@secondoftwo
 \fi
}%
\providecommand \@ifx [1]{%
 \ifx #1\expandafter \@firstoftwo
 \else \expandafter \@secondoftwo
 \fi
}%
\providecommand \natexlab [1]{#1}%
\providecommand \enquote  [1]{``#1''}%
\providecommand \bibnamefont  [1]{#1}%
\providecommand \bibfnamefont [1]{#1}%
\providecommand \citenamefont [1]{#1}%
\providecommand \href@noop [0]{\@secondoftwo}%
\providecommand \href [0]{\begingroup \@sanitize@url \@href}%
\providecommand \@href[1]{\@@startlink{#1}\@@href}%
\providecommand \@@href[1]{\endgroup#1\@@endlink}%
\providecommand \@sanitize@url [0]{\catcode `\\12\catcode `\$12\catcode
  `\&12\catcode `\#12\catcode `\^12\catcode `\_12\catcode `\%12\relax}%
\providecommand \@@startlink[1]{}%
\providecommand \@@endlink[0]{}%
\providecommand \url  [0]{\begingroup\@sanitize@url \@url }%
\providecommand \@url [1]{\endgroup\@href {#1}{\urlprefix }}%
\providecommand \urlprefix  [0]{URL }%
\providecommand \Eprint [0]{\href }%
\providecommand \doibase [0]{http://dx.doi.org/}%
\providecommand \selectlanguage [0]{\@gobble}%
\providecommand \bibinfo  [0]{\@secondoftwo}%
\providecommand \bibfield  [0]{\@secondoftwo}%
\providecommand \translation [1]{[#1]}%
\providecommand \BibitemOpen [0]{}%
\providecommand \bibitemStop [0]{}%
\providecommand \bibitemNoStop [0]{.\EOS\space}%
\providecommand \EOS [0]{\spacefactor3000\relax}%
\providecommand \BibitemShut  [1]{\csname bibitem#1\endcsname}%
\let\auto@bib@innerbib\@empty
\bibitem [{\citenamefont {Greiner}\ \emph {et~al.}(2002)\citenamefont
  {Greiner}, \citenamefont {Mandel}, \citenamefont {Esslinger}, \citenamefont
  {H\"{a}nsch},\ and\ \citenamefont {Bloch}}]{Greiner}%
  \BibitemOpen
  \bibfield  {author} {\bibinfo {author} {\bibfnamefont {M.}~\bibnamefont
  {Greiner}}, \bibinfo {author} {\bibfnamefont {O.}~\bibnamefont {Mandel}},
  \bibinfo {author} {\bibfnamefont {T.}~\bibnamefont {Esslinger}}, \bibinfo
  {author} {\bibfnamefont {T.~W.}\ \bibnamefont {H\"{a}nsch}}, \ and\ \bibinfo
  {author} {\bibfnamefont {I.}~\bibnamefont {Bloch}},\ }\bibfield  {title}
  {\enquote {\bibinfo {title} {Quantum phase transition from a superfluid to a
  mott insulator in a gas of ultracold atoms},}\ }\href@noop {} {\bibfield
  {journal} {\bibinfo  {journal} {Nature}\ }\textbf {\bibinfo {volume} {415}},\
  \bibinfo {pages} {39} (\bibinfo {year} {2002})}\BibitemShut {NoStop}%
\bibitem [{\citenamefont {Bloch}\ \emph {et~al.}(2008)\citenamefont {Bloch},
  \citenamefont {Dalibard},\ and\ \citenamefont {Zwerger}}]{Bloch}%
  \BibitemOpen
  \bibfield  {author} {\bibinfo {author} {\bibfnamefont {I.}~\bibnamefont
  {Bloch}}, \bibinfo {author} {\bibfnamefont {J.}~\bibnamefont {Dalibard}}, \
  and\ \bibinfo {author} {\bibfnamefont {W.}~\bibnamefont {Zwerger}},\
  }\bibfield  {title} {\enquote {\bibinfo {title} {Many-body physics with
  ultracold gases},}\ }\href {\doibase 10.1103/RevModPhys.80.885} {\bibfield
  {journal} {\bibinfo  {journal} {Rev. Mod. Phys.}\ }\textbf {\bibinfo {volume}
  {80}},\ \bibinfo {pages} {885} (\bibinfo {year} {2008})}\BibitemShut
  {NoStop}%
\bibitem [{\citenamefont {Lin}\ \emph {et~al.}(2011)\citenamefont {Lin},
  \citenamefont {Jim\'{e}nez-Garc\'{i}a},\ and\ \citenamefont
  {Spielman}}]{Lin1}%
  \BibitemOpen
  \bibfield  {author} {\bibinfo {author} {\bibfnamefont {Y.-J.}\ \bibnamefont
  {Lin}}, \bibinfo {author} {\bibfnamefont {K.}~\bibnamefont
  {Jim\'{e}nez-Garc\'{i}a}}, \ and\ \bibinfo {author} {\bibfnamefont {I.~B.}\
  \bibnamefont {Spielman}},\ }\bibfield  {title} {\enquote {\bibinfo {title}
  {Spin–orbit-coupled bose–einstein condensates},}\ }\href@noop {}
  {\bibfield  {journal} {\bibinfo  {journal} {Nature}\ }\textbf {\bibinfo
  {volume} {471}},\ \bibinfo {pages} {83} (\bibinfo {year} {2011})}\BibitemShut
  {NoStop}%
\bibitem [{\citenamefont {Dalibard}\ \emph {et~al.}(2011)\citenamefont
  {Dalibard}, \citenamefont {Gerbier}, \citenamefont
  {Juzeli\ifmmode~\bar{u}\else \={u}\fi{}nas},\ and\ \citenamefont
  {\"Ohberg}}]{Dalibard}%
  \BibitemOpen
  \bibfield  {author} {\bibinfo {author} {\bibfnamefont {J.}~\bibnamefont
  {Dalibard}}, \bibinfo {author} {\bibfnamefont {F.}~\bibnamefont {Gerbier}},
  \bibinfo {author} {\bibfnamefont {G.}~\bibnamefont
  {Juzeli\ifmmode~\bar{u}\else \={u}\fi{}nas}}, \ and\ \bibinfo {author}
  {\bibfnamefont {P.}~\bibnamefont {\"Ohberg}},\ }\bibfield  {title} {\enquote
  {\bibinfo {title} {Colloquium: Artificial gauge potentials for neutral
  atoms},}\ }\href {\doibase 10.1103/RevModPhys.83.1523} {\bibfield  {journal}
  {\bibinfo  {journal} {Rev. Mod. Phys.}\ }\textbf {\bibinfo {volume} {83}},\
  \bibinfo {pages} {1523} (\bibinfo {year} {2011})}\BibitemShut {NoStop}%
\bibitem [{\citenamefont {Goldman}\ \emph {et~al.}(2014)\citenamefont
  {Goldman}, \citenamefont {Juzeli{\={u}}nas}, \citenamefont {Öhberg},\ and\
  \citenamefont {Spielman}}]{Goldman}%
  \BibitemOpen
  \bibfield  {author} {\bibinfo {author} {\bibfnamefont {N.}~\bibnamefont
  {Goldman}}, \bibinfo {author} {\bibfnamefont {G.}~\bibnamefont
  {Juzeli{\={u}}nas}}, \bibinfo {author} {\bibfnamefont {P.}~\bibnamefont
  {Öhberg}}, \ and\ \bibinfo {author} {\bibfnamefont {I.~B.}\ \bibnamefont
  {Spielman}},\ }\bibfield  {title} {\enquote {\bibinfo {title} {Light-induced
  gauge fields for ultracold atoms},}\ }\href {\doibase
  10.1088/0034-4885/77/12/126401} {\bibfield  {journal} {\bibinfo  {journal}
  {Reports on Progress in Physics}\ }\textbf {\bibinfo {volume} {77}},\
  \bibinfo {pages} {126401} (\bibinfo {year} {2014})}\BibitemShut {NoStop}%
\bibitem [{\citenamefont {Matthews}\ \emph {et~al.}(1999)\citenamefont
  {Matthews}, \citenamefont {Anderson}, \citenamefont {Haljan}, \citenamefont
  {Hall}, \citenamefont {Wieman},\ and\ \citenamefont {Cornell}}]{Matthews}%
  \BibitemOpen
  \bibfield  {author} {\bibinfo {author} {\bibfnamefont {M.~R.}\ \bibnamefont
  {Matthews}}, \bibinfo {author} {\bibfnamefont {B.~P.}\ \bibnamefont
  {Anderson}}, \bibinfo {author} {\bibfnamefont {P.~C.}\ \bibnamefont
  {Haljan}}, \bibinfo {author} {\bibfnamefont {D.~S.}\ \bibnamefont {Hall}},
  \bibinfo {author} {\bibfnamefont {C.~E.}\ \bibnamefont {Wieman}}, \ and\
  \bibinfo {author} {\bibfnamefont {E.~A.}\ \bibnamefont {Cornell}},\
  }\bibfield  {title} {\enquote {\bibinfo {title} {Vortices in a bose-einstein
  condensate},}\ }\href {\doibase 10.1103/PhysRevLett.83.2498} {\bibfield
  {journal} {\bibinfo  {journal} {Phys. Rev. Lett.}\ }\textbf {\bibinfo
  {volume} {83}},\ \bibinfo {pages} {2498} (\bibinfo {year}
  {1999})}\BibitemShut {NoStop}%
\bibitem [{\citenamefont {Madison}\ \emph {et~al.}(2000)\citenamefont
  {Madison}, \citenamefont {Chevy}, \citenamefont {Wohlleben},\ and\
  \citenamefont {Dalibard}}]{Madison}%
  \BibitemOpen
  \bibfield  {author} {\bibinfo {author} {\bibfnamefont {K.~W.}\ \bibnamefont
  {Madison}}, \bibinfo {author} {\bibfnamefont {F.}~\bibnamefont {Chevy}},
  \bibinfo {author} {\bibfnamefont {W.}~\bibnamefont {Wohlleben}}, \ and\
  \bibinfo {author} {\bibfnamefont {J.}~\bibnamefont {Dalibard}},\ }\bibfield
  {title} {\enquote {\bibinfo {title} {Vortex formation in a stirred
  bose-einstein condensate},}\ }\href {\doibase 10.1103/PhysRevLett.84.806}
  {\bibfield  {journal} {\bibinfo  {journal} {Phys. Rev. Lett.}\ }\textbf
  {\bibinfo {volume} {84}},\ \bibinfo {pages} {806} (\bibinfo {year}
  {2000})}\BibitemShut {NoStop}%
\bibitem [{\citenamefont {Juzeli\ifmmode~\bar{u}\else \={u}\fi{}nas}\ and\
  \citenamefont {\"Ohberg}(2004)}]{Juzeliuas1}%
  \BibitemOpen
  \bibfield  {author} {\bibinfo {author} {\bibfnamefont {G.}~\bibnamefont
  {Juzeli\ifmmode~\bar{u}\else \={u}\fi{}nas}}\ and\ \bibinfo {author}
  {\bibfnamefont {P.}~\bibnamefont {\"Ohberg}},\ }\bibfield  {title} {\enquote
  {\bibinfo {title} {Slow light in degenerate fermi gases},}\ }\href {\doibase
  10.1103/PhysRevLett.93.033602} {\bibfield  {journal} {\bibinfo  {journal}
  {Phys. Rev. Lett.}\ }\textbf {\bibinfo {volume} {93}},\ \bibinfo {pages}
  {033602} (\bibinfo {year} {2004})}\BibitemShut {NoStop}%
\bibitem [{\citenamefont {Juzeli\ifmmode~\bar{u}\else \={u}\fi{}nas}\ \emph
  {et~al.}(2006)\citenamefont {Juzeli\ifmmode~\bar{u}\else \={u}\fi{}nas},
  \citenamefont {Ruseckas}, \citenamefont {\"Ohberg},\ and\ \citenamefont
  {Fleischhauer}}]{Juzeliuas2}%
  \BibitemOpen
  \bibfield  {author} {\bibinfo {author} {\bibfnamefont {G.}~\bibnamefont
  {Juzeli\ifmmode~\bar{u}\else \={u}\fi{}nas}}, \bibinfo {author}
  {\bibfnamefont {J.}~\bibnamefont {Ruseckas}}, \bibinfo {author}
  {\bibfnamefont {P.}~\bibnamefont {\"Ohberg}}, \ and\ \bibinfo {author}
  {\bibfnamefont {M.}~\bibnamefont {Fleischhauer}},\ }\bibfield  {title}
  {\enquote {\bibinfo {title} {Light-induced effective magnetic fields for
  ultracold atoms in planar geometries},}\ }\href {\doibase
  10.1103/PhysRevA.73.025602} {\bibfield  {journal} {\bibinfo  {journal} {Phys.
  Rev. A}\ }\textbf {\bibinfo {volume} {73}},\ \bibinfo {pages} {025602}
  (\bibinfo {year} {2006})}\BibitemShut {NoStop}%
\bibitem [{\citenamefont {Lin}\ \emph {et~al.}(2009)\citenamefont {Lin},
  \citenamefont {Compton}, \citenamefont {Jim\'{e}nez-Garc\'{i}a},
  \citenamefont {Porto},\ and\ \citenamefont {Spielman}}]{Lin}%
  \BibitemOpen
  \bibfield  {author} {\bibinfo {author} {\bibfnamefont {Y.-J.}\ \bibnamefont
  {Lin}}, \bibinfo {author} {\bibfnamefont {R.~L.}\ \bibnamefont {Compton}},
  \bibinfo {author} {\bibfnamefont {K.}~\bibnamefont {Jim\'{e}nez-Garc\'{i}a}},
  \bibinfo {author} {\bibfnamefont {J.~V.}\ \bibnamefont {Porto}}, \ and\
  \bibinfo {author} {\bibfnamefont {I.~B.}\ \bibnamefont {Spielman}},\
  }\bibfield  {title} {\enquote {\bibinfo {title} {Synthetic magnetic fields
  for ultracold neutral atoms},}\ }\href@noop {} {\bibfield  {journal}
  {\bibinfo  {journal} {Nature}\ }\textbf {\bibinfo {volume} {462}},\ \bibinfo
  {pages} {628} (\bibinfo {year} {2009})}\BibitemShut {NoStop}%
\bibitem [{\citenamefont {Miyake}\ \emph {et~al.}(2013)\citenamefont {Miyake},
  \citenamefont {Siviloglou}, \citenamefont {Kennedy}, \citenamefont {Burton},\
  and\ \citenamefont {Ketterle}}]{Miyake2013}%
  \BibitemOpen
  \bibfield  {author} {\bibinfo {author} {\bibfnamefont {H.}~\bibnamefont
  {Miyake}}, \bibinfo {author} {\bibfnamefont {G.~A.}\ \bibnamefont
  {Siviloglou}}, \bibinfo {author} {\bibfnamefont {C.~J.}\ \bibnamefont
  {Kennedy}}, \bibinfo {author} {\bibfnamefont {W.~C.}\ \bibnamefont {Burton}},
  \ and\ \bibinfo {author} {\bibfnamefont {W.}~\bibnamefont {Ketterle}},\
  }\bibfield  {title} {\enquote {\bibinfo {title} {Realizing the harper
  hamiltonian with laser-assisted tunneling in optical lattices},}\ }\href
  {\doibase 10.1103/PhysRevLett.111.185302} {\bibfield  {journal} {\bibinfo
  {journal} {Phys. Rev. Lett.}\ }\textbf {\bibinfo {volume} {111}},\ \bibinfo
  {pages} {185302} (\bibinfo {year} {2013})}\BibitemShut {NoStop}%
\bibitem [{\citenamefont {Aidelsburger}\ \emph {et~al.}(2011)\citenamefont
  {Aidelsburger}, \citenamefont {Atala}, \citenamefont {Nascimb\`ene},
  \citenamefont {Trotzky}, \citenamefont {Chen},\ and\ \citenamefont
  {Bloch}}]{Aidelsburger}%
  \BibitemOpen
  \bibfield  {author} {\bibinfo {author} {\bibfnamefont {M.}~\bibnamefont
  {Aidelsburger}}, \bibinfo {author} {\bibfnamefont {M.}~\bibnamefont {Atala}},
  \bibinfo {author} {\bibfnamefont {S.}~\bibnamefont {Nascimb\`ene}}, \bibinfo
  {author} {\bibfnamefont {S.}~\bibnamefont {Trotzky}}, \bibinfo {author}
  {\bibfnamefont {Y.-A.}\ \bibnamefont {Chen}}, \ and\ \bibinfo {author}
  {\bibfnamefont {I.}~\bibnamefont {Bloch}},\ }\bibfield  {title} {\enquote
  {\bibinfo {title} {Experimental realization of strong effective magnetic
  fields in an optical lattice},}\ }\href {\doibase
  10.1103/PhysRevLett.107.255301} {\bibfield  {journal} {\bibinfo  {journal}
  {Phys. Rev. Lett.}\ }\textbf {\bibinfo {volume} {107}},\ \bibinfo {pages}
  {255301} (\bibinfo {year} {2011})}\BibitemShut {NoStop}%
\bibitem [{\citenamefont {Parker}\ \emph {et~al.}(2013)\citenamefont {Parker},
  \citenamefont {Ha},\ and\ \citenamefont {Chin}}]{Parker2013}%
  \BibitemOpen
  \bibfield  {author} {\bibinfo {author} {\bibfnamefont {C.~V.}\ \bibnamefont
  {Parker}}, \bibinfo {author} {\bibfnamefont {L.-C.}\ \bibnamefont {Ha}}, \
  and\ \bibinfo {author} {\bibfnamefont {C.}~\bibnamefont {Chin}},\ }\bibfield
  {title} {\enquote {\bibinfo {title} {Direct observation of effective
  ferromagnetic domains of cold atoms in a shaken optical lattice},}\ }\href
  {\doibase 10.1038/nphys2789} {\bibfield  {journal} {\bibinfo  {journal}
  {Nature Physics}\ }\textbf {\bibinfo {volume} {9}},\ \bibinfo {pages} {769}
  (\bibinfo {year} {2013})}\BibitemShut {NoStop}%
\bibitem [{\citenamefont {Banerjee}\ \emph {et~al.}(2012)\citenamefont
  {Banerjee}, \citenamefont {Dalmonte}, \citenamefont {M\"uller}, \citenamefont
  {Rico}, \citenamefont {Stebler}, \citenamefont {Wiese},\ and\ \citenamefont
  {Zoller}}]{Banerjee}%
  \BibitemOpen
  \bibfield  {author} {\bibinfo {author} {\bibfnamefont {D.}~\bibnamefont
  {Banerjee}}, \bibinfo {author} {\bibfnamefont {M.}~\bibnamefont {Dalmonte}},
  \bibinfo {author} {\bibfnamefont {M.}~\bibnamefont {M\"uller}}, \bibinfo
  {author} {\bibfnamefont {E.}~\bibnamefont {Rico}}, \bibinfo {author}
  {\bibfnamefont {P.}~\bibnamefont {Stebler}}, \bibinfo {author} {\bibfnamefont
  {U.-J.}\ \bibnamefont {Wiese}}, \ and\ \bibinfo {author} {\bibfnamefont
  {P.}~\bibnamefont {Zoller}},\ }\bibfield  {title} {\enquote {\bibinfo {title}
  {Atomic quantum simulation of dynamical gauge fields coupled to fermionic
  matter: From string breaking to evolution after a quench},}\ }\href {\doibase
  10.1103/PhysRevLett.109.175302} {\bibfield  {journal} {\bibinfo  {journal}
  {Phys. Rev. Lett.}\ }\textbf {\bibinfo {volume} {109}},\ \bibinfo {pages}
  {175302} (\bibinfo {year} {2012})}\BibitemShut {NoStop}%
\bibitem [{\citenamefont {Zohar}\ \emph {et~al.}(2013)\citenamefont {Zohar},
  \citenamefont {Cirac},\ and\ \citenamefont {Reznik}}]{Zohar}%
  \BibitemOpen
  \bibfield  {author} {\bibinfo {author} {\bibfnamefont {E.}~\bibnamefont
  {Zohar}}, \bibinfo {author} {\bibfnamefont {J.~I.}\ \bibnamefont {Cirac}}, \
  and\ \bibinfo {author} {\bibfnamefont {B.}~\bibnamefont {Reznik}},\
  }\bibfield  {title} {\enquote {\bibinfo {title} {Simulating
  ($2+1$)-dimensional lattice qed with dynamical matter using ultracold
  atoms},}\ }\href {\doibase 10.1103/PhysRevLett.110.055302} {\bibfield
  {journal} {\bibinfo  {journal} {Phys. Rev. Lett.}\ }\textbf {\bibinfo
  {volume} {110}},\ \bibinfo {pages} {055302} (\bibinfo {year}
  {2013})}\BibitemShut {NoStop}%
\bibitem [{\citenamefont {Tagliacozzo}\ \emph {et~al.}(2013)\citenamefont
  {Tagliacozzo}, \citenamefont {Celi}, \citenamefont {Orland}, \citenamefont
  {Mitchell},\ and\ \citenamefont {Lewenstein}}]{Tagliacozzo}%
  \BibitemOpen
  \bibfield  {author} {\bibinfo {author} {\bibfnamefont {L.}~\bibnamefont
  {Tagliacozzo}}, \bibinfo {author} {\bibfnamefont {A.}~\bibnamefont {Celi}},
  \bibinfo {author} {\bibfnamefont {P.}~\bibnamefont {Orland}}, \bibinfo
  {author} {\bibfnamefont {M.~W.}\ \bibnamefont {Mitchell}}, \ and\ \bibinfo
  {author} {\bibfnamefont {M.}~\bibnamefont {Lewenstein}},\ }\bibfield  {title}
  {\enquote {\bibinfo {title} {Synthetic magnetic fields for ultracold neutral
  atoms},}\ }\href@noop {} {\bibfield  {journal} {\bibinfo  {journal} {Nature
  Communications}\ }\textbf {\bibinfo {volume} {4}},\ \bibinfo {pages} {2615}
  (\bibinfo {year} {2013})}\BibitemShut {NoStop}%
\bibitem [{\citenamefont {Greschner}\ \emph {et~al.}(2014)\citenamefont
  {Greschner}, \citenamefont {Sun}, \citenamefont {Poletti},\ and\
  \citenamefont {Santos}}]{Greschner}%
  \BibitemOpen
  \bibfield  {author} {\bibinfo {author} {\bibfnamefont {S.}~\bibnamefont
  {Greschner}}, \bibinfo {author} {\bibfnamefont {G.}~\bibnamefont {Sun}},
  \bibinfo {author} {\bibfnamefont {D.}~\bibnamefont {Poletti}}, \ and\
  \bibinfo {author} {\bibfnamefont {L.}~\bibnamefont {Santos}},\ }\bibfield
  {title} {\enquote {\bibinfo {title} {Density-dependent synthetic gauge fields
  using periodically modulated interactions},}\ }\href {\doibase
  10.1103/PhysRevLett.113.215303} {\bibfield  {journal} {\bibinfo  {journal}
  {Phys. Rev. Lett.}\ }\textbf {\bibinfo {volume} {113}},\ \bibinfo {pages}
  {215303} (\bibinfo {year} {2014})}\BibitemShut {NoStop}%
\bibitem [{\citenamefont {Dong}\ \emph {et~al.}(2014)\citenamefont {Dong},
  \citenamefont {Zhou}, \citenamefont {Wu}, \citenamefont {Ramachandhran},\
  and\ \citenamefont {Pu}}]{Dong}%
  \BibitemOpen
  \bibfield  {author} {\bibinfo {author} {\bibfnamefont {L.}~\bibnamefont
  {Dong}}, \bibinfo {author} {\bibfnamefont {L.}~\bibnamefont {Zhou}}, \bibinfo
  {author} {\bibfnamefont {B.}~\bibnamefont {Wu}}, \bibinfo {author}
  {\bibfnamefont {B.}~\bibnamefont {Ramachandhran}}, \ and\ \bibinfo {author}
  {\bibfnamefont {H.}~\bibnamefont {Pu}},\ }\bibfield  {title} {\enquote
  {\bibinfo {title} {Cavity-assisted dynamical spin-orbit coupling in cold
  atoms},}\ }\href {\doibase 10.1103/PhysRevA.89.011602} {\bibfield  {journal}
  {\bibinfo  {journal} {Phys. Rev. A}\ }\textbf {\bibinfo {volume} {89}},\
  \bibinfo {pages} {011602(R)} (\bibinfo {year} {2014})}\BibitemShut {NoStop}%
\bibitem [{\citenamefont {Ballantine}\ \emph {et~al.}(2017)\citenamefont
  {Ballantine}, \citenamefont {Lev},\ and\ \citenamefont
  {Keeling}}]{Ballantine}%
  \BibitemOpen
  \bibfield  {author} {\bibinfo {author} {\bibfnamefont {K.~E.}\ \bibnamefont
  {Ballantine}}, \bibinfo {author} {\bibfnamefont {B.~L.}\ \bibnamefont {Lev}},
  \ and\ \bibinfo {author} {\bibfnamefont {J.}~\bibnamefont {Keeling}},\
  }\bibfield  {title} {\enquote {\bibinfo {title} {Meissner-like effect for a
  synthetic gauge field in multimode cavity qed},}\ }\href {\doibase
  10.1103/PhysRevLett.118.045302} {\bibfield  {journal} {\bibinfo  {journal}
  {Phys. Rev. Lett.}\ }\textbf {\bibinfo {volume} {118}},\ \bibinfo {pages}
  {045302} (\bibinfo {year} {2017})}\BibitemShut {NoStop}%
\bibitem [{\citenamefont {Edmonds}\ \emph {et~al.}(2013)\citenamefont
  {Edmonds}, \citenamefont {Valiente}, \citenamefont
  {Juzeli\ifmmode~\bar{u}\else \={u}\fi{}nas}, \citenamefont {Santos},\ and\
  \citenamefont {\"Ohberg}}]{Edmonds1}%
  \BibitemOpen
  \bibfield  {author} {\bibinfo {author} {\bibfnamefont {M.~J.}\ \bibnamefont
  {Edmonds}}, \bibinfo {author} {\bibfnamefont {M.}~\bibnamefont {Valiente}},
  \bibinfo {author} {\bibfnamefont {G.}~\bibnamefont
  {Juzeli\ifmmode~\bar{u}\else \={u}\fi{}nas}}, \bibinfo {author}
  {\bibfnamefont {L.}~\bibnamefont {Santos}}, \ and\ \bibinfo {author}
  {\bibfnamefont {P.}~\bibnamefont {\"Ohberg}},\ }\bibfield  {title} {\enquote
  {\bibinfo {title} {Simulating an interacting gauge theory with ultracold bose
  gases},}\ }\href {\doibase 10.1103/PhysRevLett.110.085301} {\bibfield
  {journal} {\bibinfo  {journal} {Phys. Rev. Lett.}\ }\textbf {\bibinfo
  {volume} {110}},\ \bibinfo {pages} {085301} (\bibinfo {year}
  {2013})}\BibitemShut {NoStop}%
\bibitem [{\citenamefont {Zheng}\ \emph {et~al.}(2015)\citenamefont {Zheng},
  \citenamefont {Xiong}, \citenamefont {Juzeli\ifmmode~\bar{u}\else
  \={u}\fi{}nas},\ and\ \citenamefont {Wang}}]{Zheng}%
  \BibitemOpen
  \bibfield  {author} {\bibinfo {author} {\bibfnamefont {J.-h.}\ \bibnamefont
  {Zheng}}, \bibinfo {author} {\bibfnamefont {B.}~\bibnamefont {Xiong}},
  \bibinfo {author} {\bibfnamefont {G.}~\bibnamefont
  {Juzeli\ifmmode~\bar{u}\else \={u}\fi{}nas}}, \ and\ \bibinfo {author}
  {\bibfnamefont {D.-W.}\ \bibnamefont {Wang}},\ }\bibfield  {title} {\enquote
  {\bibinfo {title} {Topological condensate in an interaction-induced gauge
  potential},}\ }\href {\doibase 10.1103/PhysRevA.92.013604} {\bibfield
  {journal} {\bibinfo  {journal} {Phys. Rev. A}\ }\textbf {\bibinfo {volume}
  {92}},\ \bibinfo {pages} {013604} (\bibinfo {year} {2015})}\BibitemShut
  {NoStop}%
\bibitem [{\citenamefont {Martinez}\ \emph {et~al.}(2016)\citenamefont
  {Martinez}, \citenamefont {Muschik}, \citenamefont {Schindler}, \citenamefont
  {Nigg}, \citenamefont {Erhard}, \citenamefont {Heyl}, \citenamefont {Hauke},
  \citenamefont {Dalmonte}, \citenamefont {Monz}, \citenamefont {Zoller},\ and\
  \citenamefont {Blatt}}]{Martinez2016}%
  \BibitemOpen
  \bibfield  {author} {\bibinfo {author} {\bibfnamefont {E.~A.}\ \bibnamefont
  {Martinez}}, \bibinfo {author} {\bibfnamefont {C.~A.}\ \bibnamefont
  {Muschik}}, \bibinfo {author} {\bibfnamefont {P.}~\bibnamefont {Schindler}},
  \bibinfo {author} {\bibfnamefont {D.}~\bibnamefont {Nigg}}, \bibinfo {author}
  {\bibfnamefont {A.}~\bibnamefont {Erhard}}, \bibinfo {author} {\bibfnamefont
  {M.}~\bibnamefont {Heyl}}, \bibinfo {author} {\bibfnamefont {P.}~\bibnamefont
  {Hauke}}, \bibinfo {author} {\bibfnamefont {M.}~\bibnamefont {Dalmonte}},
  \bibinfo {author} {\bibfnamefont {T.}~\bibnamefont {Monz}}, \bibinfo {author}
  {\bibfnamefont {P.}~\bibnamefont {Zoller}}, \ and\ \bibinfo {author}
  {\bibfnamefont {R.}~\bibnamefont {Blatt}},\ }\bibfield  {title} {\enquote
  {\bibinfo {title} {Real-time dynamics of lattice gauge theories with a
  few-qubit quantum computer},}\ }\href {\doibase 10.1038/nature18318}
  {\bibfield  {journal} {\bibinfo  {journal} {Nature}\ }\textbf {\bibinfo
  {volume} {534}},\ \bibinfo {pages} {516} (\bibinfo {year}
  {2016})}\BibitemShut {NoStop}%
\bibitem [{\citenamefont {Clark}\ \emph {et~al.}(2018)\citenamefont {Clark},
  \citenamefont {Anderson}, \citenamefont {Feng}, \citenamefont {Gaj},
  \citenamefont {Levin},\ and\ \citenamefont {Chin}}]{Clark2018}%
  \BibitemOpen
  \bibfield  {author} {\bibinfo {author} {\bibfnamefont {L.~W.}\ \bibnamefont
  {Clark}}, \bibinfo {author} {\bibfnamefont {B.~M.}\ \bibnamefont {Anderson}},
  \bibinfo {author} {\bibfnamefont {L.}~\bibnamefont {Feng}}, \bibinfo {author}
  {\bibfnamefont {A.}~\bibnamefont {Gaj}}, \bibinfo {author} {\bibfnamefont
  {K.}~\bibnamefont {Levin}}, \ and\ \bibinfo {author} {\bibfnamefont
  {C.}~\bibnamefont {Chin}},\ }\bibfield  {title} {\enquote {\bibinfo {title}
  {Observation of density-dependent gauge fields in a bose-einstein condensate
  based on micromotion control in a shaken two-dimensional lattice},}\ }\href
  {\doibase 10.1103/PhysRevLett.121.030402} {\bibfield  {journal} {\bibinfo
  {journal} {Phys. Rev. Lett.}\ }\textbf {\bibinfo {volume} {121}},\ \bibinfo
  {pages} {030402} (\bibinfo {year} {2018})}\BibitemShut {NoStop}%
\bibitem [{\citenamefont {G{\"o}rg}\ \emph {et~al.}(2019)\citenamefont
  {G{\"o}rg}, \citenamefont {Sandholzer}, \citenamefont {Minguzzi},
  \citenamefont {Desbuquois}, \citenamefont {Messer},\ and\ \citenamefont
  {Esslinger}}]{Gorg2019}%
  \BibitemOpen
  \bibfield  {author} {\bibinfo {author} {\bibfnamefont {F.}~\bibnamefont
  {G{\"o}rg}}, \bibinfo {author} {\bibfnamefont {K.}~\bibnamefont
  {Sandholzer}}, \bibinfo {author} {\bibfnamefont {J.}~\bibnamefont
  {Minguzzi}}, \bibinfo {author} {\bibfnamefont {R.}~\bibnamefont
  {Desbuquois}}, \bibinfo {author} {\bibfnamefont {M.}~\bibnamefont {Messer}},
  \ and\ \bibinfo {author} {\bibfnamefont {T.}~\bibnamefont {Esslinger}},\
  }\bibfield  {title} {\enquote {\bibinfo {title} {Realization of
  density-dependent peierls phases to engineer quantized gauge fields coupled
  to ultracold matter},}\ }\href {\doibase 10.1038/s41567-019-0615-4}
  {\bibfield  {journal} {\bibinfo  {journal} {Nature Physics}\ }\textbf
  {\bibinfo {volume} {15}},\ \bibinfo {pages} {1161} (\bibinfo {year}
  {2019})}\BibitemShut {NoStop}%
\bibitem [{\citenamefont {Keilmann}\ \emph {et~al.}(2011)\citenamefont
  {Keilmann}, \citenamefont {Lanzmich}, \citenamefont {McCulloch},\ and\
  \citenamefont {Roncaglia}}]{Keilmann}%
  \BibitemOpen
  \bibfield  {author} {\bibinfo {author} {\bibfnamefont {T.}~\bibnamefont
  {Keilmann}}, \bibinfo {author} {\bibfnamefont {S.}~\bibnamefont {Lanzmich}},
  \bibinfo {author} {\bibfnamefont {I.}~\bibnamefont {McCulloch}}, \ and\
  \bibinfo {author} {\bibfnamefont {M.}~\bibnamefont {Roncaglia}},\ }\bibfield
  {title} {\enquote {\bibinfo {title} {Statistically induced phase transitions
  and anyons in 1d optical lattices},}\ }\href@noop {} {\bibfield  {journal}
  {\bibinfo  {journal} {Nature Communications}\ }\textbf {\bibinfo {volume}
  {2}},\ \bibinfo {pages} {361} (\bibinfo {year} {2011})}\BibitemShut {NoStop}%
\bibitem [{\citenamefont {Aglietti}\ \emph {et~al.}(1996)\citenamefont
  {Aglietti}, \citenamefont {Griguolo}, \citenamefont {Jackiw}, \citenamefont
  {Pi},\ and\ \citenamefont {Seminara}}]{Aglietti}%
  \BibitemOpen
  \bibfield  {author} {\bibinfo {author} {\bibfnamefont {U.}~\bibnamefont
  {Aglietti}}, \bibinfo {author} {\bibfnamefont {L.}~\bibnamefont {Griguolo}},
  \bibinfo {author} {\bibfnamefont {R.}~\bibnamefont {Jackiw}}, \bibinfo
  {author} {\bibfnamefont {S.-Y.}\ \bibnamefont {Pi}}, \ and\ \bibinfo {author}
  {\bibfnamefont {D.}~\bibnamefont {Seminara}},\ }\bibfield  {title} {\enquote
  {\bibinfo {title} {Anyons and chiral solitons on a line},}\ }\href {\doibase
  10.1103/PhysRevLett.77.4406} {\bibfield  {journal} {\bibinfo  {journal}
  {Phys. Rev. Lett.}\ }\textbf {\bibinfo {volume} {77}},\ \bibinfo {pages}
  {4406} (\bibinfo {year} {1996})}\BibitemShut {NoStop}%
\bibitem [{\citenamefont {Wilczek}(2012)}]{Wilczek}%
  \BibitemOpen
  \bibfield  {author} {\bibinfo {author} {\bibfnamefont {F.}~\bibnamefont
  {Wilczek}},\ }\bibfield  {title} {\enquote {\bibinfo {title} {Quantum time
  crystals},}\ }\href {\doibase 10.1103/PhysRevLett.109.160401} {\bibfield
  {journal} {\bibinfo  {journal} {Phys. Rev. Lett.}\ }\textbf {\bibinfo
  {volume} {109}},\ \bibinfo {pages} {160401} (\bibinfo {year}
  {2012})}\BibitemShut {NoStop}%
\bibitem [{\citenamefont {\"Ohberg}\ and\ \citenamefont
  {Wright}(2019)}]{Ohberg}%
  \BibitemOpen
  \bibfield  {author} {\bibinfo {author} {\bibfnamefont {P.}~\bibnamefont
  {\"Ohberg}}\ and\ \bibinfo {author} {\bibfnamefont {E.~M.}\ \bibnamefont
  {Wright}},\ }\bibfield  {title} {\enquote {\bibinfo {title} {Quantum time
  crystals and interacting gauge theories in atomic bose-einstein
  condensates},}\ }\href {\doibase 10.1103/PhysRevLett.123.250402} {\bibfield
  {journal} {\bibinfo  {journal} {Phys. Rev. Lett.}\ }\textbf {\bibinfo
  {volume} {123}},\ \bibinfo {pages} {250402} (\bibinfo {year}
  {2019})}\BibitemShut {NoStop}%
\bibitem [{\citenamefont {Öhberg}\ and\ \citenamefont
  {Wright}(2020)}]{Ohberg2020comment}%
  \BibitemOpen
  \bibfield  {author} {\bibinfo {author} {\bibfnamefont {P.}~\bibnamefont
  {Öhberg}}\ and\ \bibinfo {author} {\bibfnamefont {E.~M.}\ \bibnamefont
  {Wright}},\ }\href@noop {} {\enquote {\bibinfo {title} {Comment on "lack of a
  genuine time crystal in a chiral soliton model" by syrwid, kosior, and
  sacha},}\ } (\bibinfo {year} {2020}),\ \Eprint
  {http://arxiv.org/abs/2008.10940} {arXiv:2008.10940 [cond-mat.quant-gas]}
  \BibitemShut {NoStop}%
\bibitem [{\citenamefont {\"Ohberg}\ and\ \citenamefont
  {Wright}(2020)}]{ohberg2020response}%
  \BibitemOpen
  \bibfield  {author} {\bibinfo {author} {\bibfnamefont {P.}~\bibnamefont
  {\"Ohberg}}\ and\ \bibinfo {author} {\bibfnamefont {E.~M.}\ \bibnamefont
  {Wright}},\ }\bibfield  {title} {\enquote {\bibinfo {title} {\"ohberg and
  wright reply:},}\ }\href {\doibase 10.1103/PhysRevLett.124.178902} {\bibfield
   {journal} {\bibinfo  {journal} {Phys. Rev. Lett.}\ }\textbf {\bibinfo
  {volume} {124}},\ \bibinfo {pages} {178902} (\bibinfo {year}
  {2020})}\BibitemShut {NoStop}%
\bibitem [{\citenamefont {Syrwid}\ \emph
  {et~al.}(2020{\natexlab{a}})\citenamefont {Syrwid}, \citenamefont {Kosior},\
  and\ \citenamefont {Sacha}}]{PhysRevResearch.2.032038}%
  \BibitemOpen
  \bibfield  {author} {\bibinfo {author} {\bibfnamefont {A.}~\bibnamefont
  {Syrwid}}, \bibinfo {author} {\bibfnamefont {A.}~\bibnamefont {Kosior}}, \
  and\ \bibinfo {author} {\bibfnamefont {K.}~\bibnamefont {Sacha}},\ }\bibfield
   {title} {\enquote {\bibinfo {title} {Lack of a genuine time crystal in a
  chiral soliton model},}\ }\href {\doibase 10.1103/PhysRevResearch.2.032038}
  {\bibfield  {journal} {\bibinfo  {journal} {Phys. Rev. Research}\ }\textbf
  {\bibinfo {volume} {2}},\ \bibinfo {pages} {032038(R)} (\bibinfo {year}
  {2020}{\natexlab{a}})}\BibitemShut {NoStop}%
\bibitem [{\citenamefont {Syrwid}\ \emph
  {et~al.}(2020{\natexlab{b}})\citenamefont {Syrwid}, \citenamefont {Kosior},\
  and\ \citenamefont {Sacha}}]{syrwid2020response}%
  \BibitemOpen
  \bibfield  {author} {\bibinfo {author} {\bibfnamefont {A.}~\bibnamefont
  {Syrwid}}, \bibinfo {author} {\bibfnamefont {A.}~\bibnamefont {Kosior}}, \
  and\ \bibinfo {author} {\bibfnamefont {K.}~\bibnamefont {Sacha}},\
  }\href@noop {} {\enquote {\bibinfo {title} {Response to comment on "lack of a
  genuine time crystal in a chiral soliton model" by \"ohberg and wright},}\ }
  (\bibinfo {year} {2020}{\natexlab{b}}),\ \Eprint
  {http://arxiv.org/abs/2010.00414} {arXiv:2010.00414 [cond-mat.quant-gas]}
  \BibitemShut {NoStop}%
\bibitem [{\citenamefont {{Edmonds, M. J.}}\ \emph {et~al.}(2015)\citenamefont
  {{Edmonds, M. J.}}, \citenamefont {{Valiente, M.}},\ and\ \citenamefont
  {{\"Ohberg, P.}}}]{Edmonds2}%
  \BibitemOpen
  \bibfield  {author} {\bibinfo {author} {\bibnamefont {{Edmonds, M. J.}}},
  \bibinfo {author} {\bibnamefont {{Valiente, M.}}}, \ and\ \bibinfo {author}
  {\bibnamefont {{\"Ohberg, P.}}},\ }\bibfield  {title} {\enquote {\bibinfo
  {title} {Elementary excitations of chiral bose-einstein condensates},}\
  }\href {\doibase 10.1209/0295-5075/110/36004} {\bibfield  {journal} {\bibinfo
   {journal} {EPL}\ }\textbf {\bibinfo {volume} {110}},\ \bibinfo {pages}
  {36004} (\bibinfo {year} {2015})}\BibitemShut {NoStop}%
\bibitem [{\citenamefont {Dingwall}\ \emph {et~al.}(2018)\citenamefont
  {Dingwall}, \citenamefont {Edmonds}, \citenamefont {Helm}, \citenamefont
  {Malomed},\ and\ \citenamefont {Öhberg}}]{Dingwall}%
  \BibitemOpen
  \bibfield  {author} {\bibinfo {author} {\bibfnamefont {R.~J.}\ \bibnamefont
  {Dingwall}}, \bibinfo {author} {\bibfnamefont {M.~J.}\ \bibnamefont
  {Edmonds}}, \bibinfo {author} {\bibfnamefont {J.~L.}\ \bibnamefont {Helm}},
  \bibinfo {author} {\bibfnamefont {B.~A.}\ \bibnamefont {Malomed}}, \ and\
  \bibinfo {author} {\bibfnamefont {P.}~\bibnamefont {Öhberg}},\ }\bibfield
  {title} {\enquote {\bibinfo {title} {Non-integrable dynamics of matter-wave
  solitons in a density-dependent gauge theory},}\ }\href {\doibase
  10.1088/1367-2630/aab29e} {\bibfield  {journal} {\bibinfo  {journal} {New
  Journal of Physics}\ }\textbf {\bibinfo {volume} {20}},\ \bibinfo {pages}
  {043004} (\bibinfo {year} {2018})}\BibitemShut {NoStop}%
\bibitem [{\citenamefont {Saleh}\ and\ \citenamefont
  {{\"O}hberg}(2018)}]{Saleh}%
  \BibitemOpen
  \bibfield  {author} {\bibinfo {author} {\bibfnamefont {M.}~\bibnamefont
  {Saleh}}\ and\ \bibinfo {author} {\bibfnamefont {P.}~\bibnamefont
  {{\"O}hberg}},\ }\bibfield  {title} {{\selectlanguage {English}\enquote
  {\bibinfo {title} {Trapped bose–einstein condensates in the presence of a
  current nonlinearity},}\ }}\href {\doibase 10.1088/1361-6455/aaa64b}
  {\bibfield  {journal} {\bibinfo  {journal} {Journal of Physics B: Atomic,
  Molecular and Optical Physics}\ }\textbf {\bibinfo {volume} {51}} (\bibinfo
  {year} {2018}),\ 10.1088/1361-6455/aaa64b}\BibitemShut {NoStop}%
\bibitem [{\citenamefont {Benjamin}\ and\ \citenamefont
  {Feir}(1967)}]{Benjamin}%
  \BibitemOpen
  \bibfield  {author} {\bibinfo {author} {\bibfnamefont {T.~B.}\ \bibnamefont
  {Benjamin}}\ and\ \bibinfo {author} {\bibfnamefont {J.~E.}\ \bibnamefont
  {Feir}},\ }\bibfield  {title} {\enquote {\bibinfo {title} {The disintegration
  of wave trains on deep water part 1. theory},}\ }\href {\doibase
  10.1017/S002211206700045X} {\bibfield  {journal} {\bibinfo  {journal}
  {Journal of Fluid Mechanics}\ }\textbf {\bibinfo {volume} {27}},\ \bibinfo
  {pages} {417–430} (\bibinfo {year} {1967})}\BibitemShut {NoStop}%
\bibitem [{\citenamefont {Agrawal}()}]{Agrawal}%
  \BibitemOpen
  \bibfield  {author} {\bibinfo {author} {\bibfnamefont {G.~P.}\ \bibnamefont
  {Agrawal}},\ }\href@noop {} {\emph {\bibinfo {title} {Nonlinear Fiber
  Optics}}}\BibitemShut {NoStop}%
\bibitem [{\citenamefont {Mithun}\ \emph {et~al.}(2020)\citenamefont {Mithun},
  \citenamefont {Maluckov}, \citenamefont {Kasamatsu}, \citenamefont
  {Malomed},\ and\ \citenamefont {Khare}}]{Mithun_2020}%
  \BibitemOpen
  \bibfield  {author} {\bibinfo {author} {\bibfnamefont {T.}~\bibnamefont
  {Mithun}}, \bibinfo {author} {\bibfnamefont {A.}~\bibnamefont {Maluckov}},
  \bibinfo {author} {\bibfnamefont {K.}~\bibnamefont {Kasamatsu}}, \bibinfo
  {author} {\bibfnamefont {B.~A.}\ \bibnamefont {Malomed}}, \ and\ \bibinfo
  {author} {\bibfnamefont {A.}~\bibnamefont {Khare}},\ }\bibfield  {title}
  {\enquote {\bibinfo {title} {Modulational instability, inter-component
  asymmetry, and formation of quantum droplets in one-dimensional binary bose
  gases},}\ }\href {\doibase 10.3390/sym12010174} {\bibfield  {journal}
  {\bibinfo  {journal} {Symmetry}\ }\textbf {\bibinfo {volume} {12}},\ \bibinfo
  {pages} {174} (\bibinfo {year} {2020})}\BibitemShut {NoStop}%
\bibitem [{\citenamefont {Everitt}\ \emph {et~al.}(2017)\citenamefont
  {Everitt}, \citenamefont {Sooriyabandara}, \citenamefont {Guasoni},
  \citenamefont {Wigley}, \citenamefont {Wei}, \citenamefont {McDonald},
  \citenamefont {Hardman}, \citenamefont {Manju}, \citenamefont {Close},
  \citenamefont {Kuhn}, \citenamefont {Szigeti}, \citenamefont {Kivshar},\ and\
  \citenamefont {Robins}}]{Everitt}%
  \BibitemOpen
  \bibfield  {author} {\bibinfo {author} {\bibfnamefont {P.~J.}\ \bibnamefont
  {Everitt}}, \bibinfo {author} {\bibfnamefont {M.~A.}\ \bibnamefont
  {Sooriyabandara}}, \bibinfo {author} {\bibfnamefont {M.}~\bibnamefont
  {Guasoni}}, \bibinfo {author} {\bibfnamefont {P.~B.}\ \bibnamefont {Wigley}},
  \bibinfo {author} {\bibfnamefont {C.~H.}\ \bibnamefont {Wei}}, \bibinfo
  {author} {\bibfnamefont {G.~D.}\ \bibnamefont {McDonald}}, \bibinfo {author}
  {\bibfnamefont {K.~S.}\ \bibnamefont {Hardman}}, \bibinfo {author}
  {\bibfnamefont {P.}~\bibnamefont {Manju}}, \bibinfo {author} {\bibfnamefont
  {J.~D.}\ \bibnamefont {Close}}, \bibinfo {author} {\bibfnamefont {C.~C.~N.}\
  \bibnamefont {Kuhn}}, \bibinfo {author} {\bibfnamefont {S.~S.}\ \bibnamefont
  {Szigeti}}, \bibinfo {author} {\bibfnamefont {Y.~S.}\ \bibnamefont
  {Kivshar}}, \ and\ \bibinfo {author} {\bibfnamefont {N.~P.}\ \bibnamefont
  {Robins}},\ }\bibfield  {title} {\enquote {\bibinfo {title} {Observation of a
  modulational instability in bose-einstein condensates},}\ }\href {\doibase
  10.1103/PhysRevA.96.041601} {\bibfield  {journal} {\bibinfo  {journal} {Phys.
  Rev. A}\ }\textbf {\bibinfo {volume} {96}},\ \bibinfo {pages} {041601(R)}
  (\bibinfo {year} {2017})}\BibitemShut {NoStop}%
\bibitem [{\citenamefont {Nguyen}\ \emph {et~al.}(2017)\citenamefont {Nguyen},
  \citenamefont {Luo},\ and\ \citenamefont {Hulet}}]{Nguyen}%
  \BibitemOpen
  \bibfield  {author} {\bibinfo {author} {\bibfnamefont {J.~H.~V.}\
  \bibnamefont {Nguyen}}, \bibinfo {author} {\bibfnamefont {D.}~\bibnamefont
  {Luo}}, \ and\ \bibinfo {author} {\bibfnamefont {R.~G.}\ \bibnamefont
  {Hulet}},\ }\bibfield  {title} {\enquote {\bibinfo {title} {Formation of
  matter-wave soliton trains by modulational instability},}\ }\href {\doibase
  10.1126/science.aal3220} {\bibfield  {journal} {\bibinfo  {journal}
  {Science}\ }\textbf {\bibinfo {volume} {356}},\ \bibinfo {pages} {422}
  (\bibinfo {year} {2017})},\ \Eprint
  {http://arxiv.org/abs/https://science.sciencemag.org/content/356/6336/422.full.pdf}
  {https://science.sciencemag.org/content/356/6336/422.full.pdf} \BibitemShut
  {NoStop}%
\bibitem [{\citenamefont {Sanz}\ \emph {et~al.}(2019)\citenamefont {Sanz},
  \citenamefont {Fr{\"o}lian}, \citenamefont {Chisholm}, \citenamefont
  {Cabrera},\ and\ \citenamefont {Tarruell}}]{sanz2019interaction}%
  \BibitemOpen
  \bibfield  {author} {\bibinfo {author} {\bibfnamefont {J.}~\bibnamefont
  {Sanz}}, \bibinfo {author} {\bibfnamefont {A.}~\bibnamefont {Fr{\"o}lian}},
  \bibinfo {author} {\bibfnamefont {C.}~\bibnamefont {Chisholm}}, \bibinfo
  {author} {\bibfnamefont {C.}~\bibnamefont {Cabrera}}, \ and\ \bibinfo
  {author} {\bibfnamefont {L.}~\bibnamefont {Tarruell}},\ }\bibfield  {title}
  {\enquote {\bibinfo {title} {Interaction control and bright solitons in
  coherently-coupled bose-einstein condensates},}\ }\href@noop {} {\bibfield
  {journal} {\bibinfo  {journal} {arXiv preprint arXiv:1912.06041}\ } (\bibinfo
  {year} {2019})}\BibitemShut {NoStop}%
\bibitem [{\citenamefont {Theocharis}\ \emph {et~al.}(2003)\citenamefont
  {Theocharis}, \citenamefont {Rapti}, \citenamefont {Kevrekidis},
  \citenamefont {Frantzeskakis},\ and\ \citenamefont {Konotop}}]{Theocharis}%
  \BibitemOpen
  \bibfield  {author} {\bibinfo {author} {\bibfnamefont {G.}~\bibnamefont
  {Theocharis}}, \bibinfo {author} {\bibfnamefont {Z.}~\bibnamefont {Rapti}},
  \bibinfo {author} {\bibfnamefont {P.~G.}\ \bibnamefont {Kevrekidis}},
  \bibinfo {author} {\bibfnamefont {D.~J.}\ \bibnamefont {Frantzeskakis}}, \
  and\ \bibinfo {author} {\bibfnamefont {V.~V.}\ \bibnamefont {Konotop}},\
  }\bibfield  {title} {\enquote {\bibinfo {title} {Modulational instability of
  gross-pitaevskii-type equations in $1+1$ dimensions},}\ }\href {\doibase
  10.1103/PhysRevA.67.063610} {\bibfield  {journal} {\bibinfo  {journal} {Phys.
  Rev. A}\ }\textbf {\bibinfo {volume} {67}},\ \bibinfo {pages} {063610}
  (\bibinfo {year} {2003})}\BibitemShut {NoStop}%
\bibitem [{\citenamefont {Salasnich}\ \emph {et~al.}(2003)\citenamefont
  {Salasnich}, \citenamefont {Parola},\ and\ \citenamefont
  {Reatto}}]{Salasnich}%
  \BibitemOpen
  \bibfield  {author} {\bibinfo {author} {\bibfnamefont {L.}~\bibnamefont
  {Salasnich}}, \bibinfo {author} {\bibfnamefont {A.}~\bibnamefont {Parola}}, \
  and\ \bibinfo {author} {\bibfnamefont {L.}~\bibnamefont {Reatto}},\
  }\bibfield  {title} {\enquote {\bibinfo {title} {Modulational instability and
  complex dynamics of confined matter-wave solitons},}\ }\href {\doibase
  10.1103/PhysRevLett.91.080405} {\bibfield  {journal} {\bibinfo  {journal}
  {Phys. Rev. Lett.}\ }\textbf {\bibinfo {volume} {91}},\ \bibinfo {pages}
  {080405} (\bibinfo {year} {2003})}\BibitemShut {NoStop}%
\bibitem [{\citenamefont {Goldstein}\ and\ \citenamefont
  {Meystre}(1997)}]{Goldstein}%
  \BibitemOpen
  \bibfield  {author} {\bibinfo {author} {\bibfnamefont {E.~V.}\ \bibnamefont
  {Goldstein}}\ and\ \bibinfo {author} {\bibfnamefont {P.}~\bibnamefont
  {Meystre}},\ }\bibfield  {title} {\enquote {\bibinfo {title} {Quasiparticle
  instabilities in multicomponent atomic condensates},}\ }\href {\doibase
  10.1103/PhysRevA.55.2935} {\bibfield  {journal} {\bibinfo  {journal} {Phys.
  Rev. A}\ }\textbf {\bibinfo {volume} {55}},\ \bibinfo {pages} {2935}
  (\bibinfo {year} {1997})}\BibitemShut {NoStop}%
\bibitem [{\citenamefont {Kasamatsu}\ and\ \citenamefont
  {Tsubota}(2004)}]{Kasamatsu1}%
  \BibitemOpen
  \bibfield  {author} {\bibinfo {author} {\bibfnamefont {K.}~\bibnamefont
  {Kasamatsu}}\ and\ \bibinfo {author} {\bibfnamefont {M.}~\bibnamefont
  {Tsubota}},\ }\bibfield  {title} {\enquote {\bibinfo {title} {Multiple domain
  formation induced by modulation instability in two-component bose-einstein
  condensates},}\ }\href {\doibase 10.1103/PhysRevLett.93.100402} {\bibfield
  {journal} {\bibinfo  {journal} {Phys. Rev. Lett.}\ }\textbf {\bibinfo
  {volume} {93}},\ \bibinfo {pages} {100402} (\bibinfo {year}
  {2004})}\BibitemShut {NoStop}%
\bibitem [{\citenamefont {Kasamatsu}\ and\ \citenamefont
  {Tsubota}(2006)}]{Kasamatsu2}%
  \BibitemOpen
  \bibfield  {author} {\bibinfo {author} {\bibfnamefont {K.}~\bibnamefont
  {Kasamatsu}}\ and\ \bibinfo {author} {\bibfnamefont {M.}~\bibnamefont
  {Tsubota}},\ }\bibfield  {title} {\enquote {\bibinfo {title} {Modulation
  instability and solitary-wave formation in two-component bose-einstein
  condensates},}\ }\href {\doibase 10.1103/PhysRevA.74.013617} {\bibfield
  {journal} {\bibinfo  {journal} {Phys. Rev. A}\ }\textbf {\bibinfo {volume}
  {74}},\ \bibinfo {pages} {013617} (\bibinfo {year} {2006})}\BibitemShut
  {NoStop}%
\bibitem [{\citenamefont {Bhat}\ \emph {et~al.}(2015)\citenamefont {Bhat},
  \citenamefont {Mithun}, \citenamefont {Malomed},\ and\ \citenamefont
  {Porsezian}}]{Ishfaq}%
  \BibitemOpen
  \bibfield  {author} {\bibinfo {author} {\bibfnamefont {I.~A.}\ \bibnamefont
  {Bhat}}, \bibinfo {author} {\bibfnamefont {T.}~\bibnamefont {Mithun}},
  \bibinfo {author} {\bibfnamefont {B.~A.}\ \bibnamefont {Malomed}}, \ and\
  \bibinfo {author} {\bibfnamefont {K.}~\bibnamefont {Porsezian}},\ }\bibfield
  {title} {\enquote {\bibinfo {title} {Modulational instability in binary
  spin-orbit-coupled bose-einstein condensates},}\ }\href {\doibase
  10.1103/PhysRevA.92.063606} {\bibfield  {journal} {\bibinfo  {journal} {Phys.
  Rev. A}\ }\textbf {\bibinfo {volume} {92}},\ \bibinfo {pages} {063606}
  (\bibinfo {year} {2015})}\BibitemShut {NoStop}%
\bibitem [{\citenamefont {Mithun}\ and\ \citenamefont
  {Kasamatsu}(2019)}]{Mithun}%
  \BibitemOpen
  \bibfield  {author} {\bibinfo {author} {\bibfnamefont {T.}~\bibnamefont
  {Mithun}}\ and\ \bibinfo {author} {\bibfnamefont {K.}~\bibnamefont
  {Kasamatsu}},\ }\bibfield  {title} {\enquote {\bibinfo {title} {Modulation
  instability associated nonlinear dynamics of spin{\textendash}orbit coupled
  bose{\textendash}einstein condensates},}\ }\href {\doibase
  10.1088/1361-6455/aafbdd} {\bibfield  {journal} {\bibinfo  {journal} {Journal
  of Physics B: Atomic, Molecular and Optical Physics}\ }\textbf {\bibinfo
  {volume} {52}},\ \bibinfo {pages} {045301} (\bibinfo {year}
  {2019})}\BibitemShut {NoStop}%
\bibitem [{\citenamefont {Li}\ \emph {et~al.}(2019)\citenamefont {Li},
  \citenamefont {Cheng}, \citenamefont {Zhang},\ and\ \citenamefont
  {Xue}}]{Li}%
  \BibitemOpen
  \bibfield  {author} {\bibinfo {author} {\bibfnamefont {X.-X.}\ \bibnamefont
  {Li}}, \bibinfo {author} {\bibfnamefont {R.-J.}\ \bibnamefont {Cheng}},
  \bibinfo {author} {\bibfnamefont {A.-X.}\ \bibnamefont {Zhang}}, \ and\
  \bibinfo {author} {\bibfnamefont {J.-K.}\ \bibnamefont {Xue}},\ }\bibfield
  {title} {\enquote {\bibinfo {title} {Modulational instability of
  bose-einstein condensates with helicoidal spin-orbit coupling},}\ }\href
  {\doibase 10.1103/PhysRevE.100.032220} {\bibfield  {journal} {\bibinfo
  {journal} {Phys. Rev. E}\ }\textbf {\bibinfo {volume} {100}},\ \bibinfo
  {pages} {032220} (\bibinfo {year} {2019})}\BibitemShut {NoStop}%
\bibitem [{\citenamefont {Jackiw}(1997)}]{Jackiw}%
  \BibitemOpen
  \bibfield  {author} {\bibinfo {author} {\bibfnamefont {R.}~\bibnamefont
  {Jackiw}},\ }\bibfield  {title} {\enquote {\bibinfo {title} {A
  nonrelativistic chiral soliton in one dimension},}\ }\href {\doibase
  10.2991/jnmp.1997.4.3-4.2} {\bibfield  {journal} {\bibinfo  {journal}
  {Journal of Nonlinear Mathematical Physics}\ }\textbf {\bibinfo {volume}
  {4}},\ \bibinfo {pages} {261} (\bibinfo {year} {1997})},\ \Eprint
  {http://arxiv.org/abs/https://doi.org/10.2991/jnmp.1997.4.3-4.2}
  {https://doi.org/10.2991/jnmp.1997.4.3-4.2} \BibitemShut {NoStop}%
\bibitem [{\citenamefont {Gordon}(1986)}]{Raman}%
  \BibitemOpen
  \bibfield  {author} {\bibinfo {author} {\bibfnamefont {J.~P.}\ \bibnamefont
  {Gordon}},\ }\bibfield  {title} {\enquote {\bibinfo {title} {Theory of the
  soliton self-frequency shift},}\ }\href {\doibase 10.1364/OL.11.000662}
  {\bibfield  {journal} {\bibinfo  {journal} {Opt. Lett.}\ }\textbf {\bibinfo
  {volume} {11}},\ \bibinfo {pages} {662} (\bibinfo {year} {1986})}\BibitemShut
  {NoStop}%
\bibitem [{\citenamefont {Gromov}\ and\ \citenamefont
  {Malomed}(2013)}]{Gromov}%
  \BibitemOpen
  \bibfield  {author} {\bibinfo {author} {\bibfnamefont {E.~M.}\ \bibnamefont
  {Gromov}}\ and\ \bibinfo {author} {\bibfnamefont {B.~A.}\ \bibnamefont
  {Malomed}},\ }\bibfield  {title} {\enquote {\bibinfo {title} {Soliton
  dynamics in an extended nonlinear schrödinger equation with a spatial
  counterpart of the stimulated raman scattering},}\ }\href {\doibase
  10.1017/S0022377813000743} {\bibfield  {journal} {\bibinfo  {journal}
  {Journal of Plasma Physics}\ }\textbf {\bibinfo {volume} {79}},\ \bibinfo
  {pages} {1057–1062} (\bibinfo {year} {2013})}\BibitemShut {NoStop}%
\bibitem [{\citenamefont {Ichikawa}(1979)}]{Landau}%
  \BibitemOpen
  \bibfield  {author} {\bibinfo {author} {\bibfnamefont {Y.~H.}\ \bibnamefont
  {Ichikawa}},\ }\bibfield  {title} {\enquote {\bibinfo {title} {Topics on
  solitons in plasmas},}\ }\href {\doibase 10.1088/0031-8949/20/3-4/002}
  {\bibfield  {journal} {\bibinfo  {journal} {Physica Scripta}\ }\textbf
  {\bibinfo {volume} {20}},\ \bibinfo {pages} {296} (\bibinfo {year}
  {1979})}\BibitemShut {NoStop}%
\bibitem [{\citenamefont {Choi}\ \emph {et~al.}(1998)\citenamefont {Choi},
  \citenamefont {Morgan},\ and\ \citenamefont {Burnett}}]{Choi}%
  \BibitemOpen
  \bibfield  {author} {\bibinfo {author} {\bibfnamefont {S.}~\bibnamefont
  {Choi}}, \bibinfo {author} {\bibfnamefont {S.~A.}\ \bibnamefont {Morgan}}, \
  and\ \bibinfo {author} {\bibfnamefont {K.}~\bibnamefont {Burnett}},\
  }\bibfield  {title} {\enquote {\bibinfo {title} {Phenomenological damping in
  trapped atomic bose-einstein condensates},}\ }\href {\doibase
  10.1103/PhysRevA.57.4057} {\bibfield  {journal} {\bibinfo  {journal} {Phys.
  Rev. A}\ }\textbf {\bibinfo {volume} {57}},\ \bibinfo {pages} {4057}
  (\bibinfo {year} {1998})}\BibitemShut {NoStop}%
\bibitem [{\citenamefont {Bogolyubov}(1947)}]{Bogolyubov}%
  \BibitemOpen
  \bibfield  {author} {\bibinfo {author} {\bibfnamefont {N.~N.}\ \bibnamefont
  {Bogolyubov}},\ }\bibfield  {title} {\enquote {\bibinfo {title} {On the
  theory of superfluidity},}\ }\href@noop {} {\bibfield  {journal} {\bibinfo
  {journal} {J. Phys.(USSR)}\ }\textbf {\bibinfo {volume} {11}},\ \bibinfo
  {pages} {23} (\bibinfo {year} {1947})},\ \bibinfo {note} {[Izv. Akad. Nauk
  Ser. Fiz.11,77(1947)]}\BibitemShut {NoStop}%
\bibitem [{\citenamefont {Blatt}\ \emph {et~al.}(2011)\citenamefont {Blatt},
  \citenamefont {Nicholson}, \citenamefont {Bloom}, \citenamefont {Williams},
  \citenamefont {Thomsen}, \citenamefont {Julienne},\ and\ \citenamefont
  {Ye}}]{Blatt}%
  \BibitemOpen
  \bibfield  {author} {\bibinfo {author} {\bibfnamefont {S.}~\bibnamefont
  {Blatt}}, \bibinfo {author} {\bibfnamefont {T.~L.}\ \bibnamefont
  {Nicholson}}, \bibinfo {author} {\bibfnamefont {B.~J.}\ \bibnamefont
  {Bloom}}, \bibinfo {author} {\bibfnamefont {J.~R.}\ \bibnamefont {Williams}},
  \bibinfo {author} {\bibfnamefont {J.~W.}\ \bibnamefont {Thomsen}}, \bibinfo
  {author} {\bibfnamefont {P.~S.}\ \bibnamefont {Julienne}}, \ and\ \bibinfo
  {author} {\bibfnamefont {J.}~\bibnamefont {Ye}},\ }\bibfield  {title}
  {\enquote {\bibinfo {title} {Measurement of optical feshbach resonances in an
  ideal gas},}\ }\href {\doibase 10.1103/PhysRevLett.107.073202} {\bibfield
  {journal} {\bibinfo  {journal} {Phys. Rev. Lett.}\ }\textbf {\bibinfo
  {volume} {107}},\ \bibinfo {pages} {073202} (\bibinfo {year}
  {2011})}\BibitemShut {NoStop}%
\bibitem [{\citenamefont {Inouye}\ \emph {et~al.}(1998)\citenamefont {Inouye},
  \citenamefont {Andrews}, \citenamefont {Stenger}, \citenamefont {Miesner},
  \citenamefont {Stamper-Kurn},\ and\ \citenamefont {Ketterle}}]{Inouye}%
  \BibitemOpen
  \bibfield  {author} {\bibinfo {author} {\bibfnamefont {S.}~\bibnamefont
  {Inouye}}, \bibinfo {author} {\bibfnamefont {M.~R.}\ \bibnamefont {Andrews}},
  \bibinfo {author} {\bibfnamefont {J.}~\bibnamefont {Stenger}}, \bibinfo
  {author} {\bibfnamefont {H.-J.}\ \bibnamefont {Miesner}}, \bibinfo {author}
  {\bibfnamefont {D.~M.}\ \bibnamefont {Stamper-Kurn}}, \ and\ \bibinfo
  {author} {\bibfnamefont {W.}~\bibnamefont {Ketterle}},\ }\bibfield  {title}
  {\enquote {\bibinfo {title} {Observation of feshbach resonances in a
  bose–einstein condensate},}\ }\href@noop {} {\bibfield  {journal} {\bibinfo
   {journal} {Nature}\ }\textbf {\bibinfo {volume} {392}},\ \bibinfo {pages}
  {151} (\bibinfo {year} {1998})}\BibitemShut {NoStop}%
\bibitem [{\citenamefont {Cornish}\ \emph {et~al.}(2000)\citenamefont
  {Cornish}, \citenamefont {Claussen}, \citenamefont {Roberts}, \citenamefont
  {Cornell},\ and\ \citenamefont {Wieman}}]{Cornish}%
  \BibitemOpen
  \bibfield  {author} {\bibinfo {author} {\bibfnamefont {S.~L.}\ \bibnamefont
  {Cornish}}, \bibinfo {author} {\bibfnamefont {N.~R.}\ \bibnamefont
  {Claussen}}, \bibinfo {author} {\bibfnamefont {J.~L.}\ \bibnamefont
  {Roberts}}, \bibinfo {author} {\bibfnamefont {E.~A.}\ \bibnamefont
  {Cornell}}, \ and\ \bibinfo {author} {\bibfnamefont {C.~E.}\ \bibnamefont
  {Wieman}},\ }\bibfield  {title} {\enquote {\bibinfo {title} {Stable
  ${}^{85}\mathrm{Rb}$ bose-einstein condensates with widely tunable
  interactions},}\ }\href {\doibase 10.1103/PhysRevLett.85.1795} {\bibfield
  {journal} {\bibinfo  {journal} {Phys. Rev. Lett.}\ }\textbf {\bibinfo
  {volume} {85}},\ \bibinfo {pages} {1795} (\bibinfo {year}
  {2000})}\BibitemShut {NoStop}%
\bibitem [{\citenamefont {Sarafyan}(1972)}]{sarafyan1972improved}%
  \BibitemOpen
  \bibfield  {author} {\bibinfo {author} {\bibfnamefont {D.}~\bibnamefont
  {Sarafyan}},\ }\bibfield  {title} {\enquote {\bibinfo {title} {Improved
  sixth-order runge-kutta formulas and approximate continuous solution of
  ordinary differential equations},}\ }\href@noop {} {\bibfield  {journal}
  {\bibinfo  {journal} {Journal of Mathematical Analysis and Applications}\
  }\textbf {\bibinfo {volume} {40}},\ \bibinfo {pages} {436} (\bibinfo {year}
  {1972})}\BibitemShut {NoStop}%
\bibitem [{\citenamefont {Edwards}\ and\ \citenamefont
  {Burnett}(1995)}]{PhysRevA.51.1382}%
  \BibitemOpen
  \bibfield  {author} {\bibinfo {author} {\bibfnamefont {M.}~\bibnamefont
  {Edwards}}\ and\ \bibinfo {author} {\bibfnamefont {K.}~\bibnamefont
  {Burnett}},\ }\bibfield  {title} {\enquote {\bibinfo {title} {Numerical
  solution of the nonlinear schr\"odinger equation for small samples of trapped
  neutral atoms},}\ }\href {\doibase 10.1103/PhysRevA.51.1382} {\bibfield
  {journal} {\bibinfo  {journal} {Phys. Rev. A}\ }\textbf {\bibinfo {volume}
  {51}},\ \bibinfo {pages} {1382} (\bibinfo {year} {1995})}\BibitemShut
  {NoStop}%
\bibitem [{\citenamefont {Kanamoto}\ \emph {et~al.}(2003)\citenamefont
  {Kanamoto}, \citenamefont {Saito},\ and\ \citenamefont {Ueda}}]{Ueda}%
  \BibitemOpen
  \bibfield  {author} {\bibinfo {author} {\bibfnamefont {R.}~\bibnamefont
  {Kanamoto}}, \bibinfo {author} {\bibfnamefont {H.}~\bibnamefont {Saito}}, \
  and\ \bibinfo {author} {\bibfnamefont {M.}~\bibnamefont {Ueda}},\ }\bibfield
  {title} {\enquote {\bibinfo {title} {Quantum phase transition in
  one-dimensional bose-einstein condensates with attractive interactions},}\
  }\href {\doibase 10.1103/PhysRevA.67.013608} {\bibfield  {journal} {\bibinfo
  {journal} {Phys. Rev. A}\ }\textbf {\bibinfo {volume} {67}},\ \bibinfo
  {pages} {013608} (\bibinfo {year} {2003})}\BibitemShut {NoStop}%
\bibitem [{\citenamefont {Abdullaev}\ \emph {et~al.}(2001)\citenamefont
  {Abdullaev}, \citenamefont {Gammal}, \citenamefont {Tomio},\ and\
  \citenamefont {Frederico}}]{quintic1}%
  \BibitemOpen
  \bibfield  {author} {\bibinfo {author} {\bibfnamefont {F.~K.}\ \bibnamefont
  {Abdullaev}}, \bibinfo {author} {\bibfnamefont {A.}~\bibnamefont {Gammal}},
  \bibinfo {author} {\bibfnamefont {L.}~\bibnamefont {Tomio}}, \ and\ \bibinfo
  {author} {\bibfnamefont {T.}~\bibnamefont {Frederico}},\ }\bibfield  {title}
  {\enquote {\bibinfo {title} {Stability of trapped bose-einstein
  condensates},}\ }\href {\doibase 10.1103/PhysRevA.63.043604} {\bibfield
  {journal} {\bibinfo  {journal} {Phys. Rev. A}\ }\textbf {\bibinfo {volume}
  {63}},\ \bibinfo {pages} {043604} (\bibinfo {year} {2001})}\BibitemShut
  {NoStop}%
\bibitem [{\citenamefont {Abdullaev}\ and\ \citenamefont
  {Salerno}(2005)}]{quintic2}%
  \BibitemOpen
  \bibfield  {author} {\bibinfo {author} {\bibfnamefont {F.~K.}\ \bibnamefont
  {Abdullaev}}\ and\ \bibinfo {author} {\bibfnamefont {M.}~\bibnamefont
  {Salerno}},\ }\bibfield  {title} {\enquote {\bibinfo {title} {Gap-townes
  solitons and localized excitations in low-dimensional bose-einstein
  condensates in optical lattices},}\ }\href
  {https://journals.aps.org/pra/abstract/10.1103/PhysRevA.72.033617} {\bibfield
   {journal} {\bibinfo  {journal} {Phys. Rev. A}\ }\textbf {\bibinfo {volume}
  {72}},\ \bibinfo {pages} {033617} (\bibinfo {year} {2005})}\BibitemShut
  {NoStop}%
\bibitem [{\citenamefont {Fibich}(2015)}]{Fibich}%
  \BibitemOpen
  \bibfield  {author} {\bibinfo {author} {\bibfnamefont {G.}~\bibnamefont
  {Fibich}},\ }\bibfield  {title} {\enquote {\bibinfo {title} {The nonlinear
  schr\"{o}dinger equation: Singular solutions and optical collapse},}\
  }\href@noop {} {\  (\bibinfo {year} {2015})}\BibitemShut {NoStop}%
\end{thebibliography}%


\providecommand{\noopsort}[1]{}\providecommand{\singleletter}[1]{#1}%
%

\end{document}